\documentclass[twocolumn]{aastex631}
\shorttitle{Broadband X-ray spectral analysis of Mrk~739}
\shortauthors{Inaba et al.}
\graphicspath{{./}{figures/}}

\begin{document}

\title{Broadband X-ray Spectral Analysis of the Dual AGN System Mrk~739}
\author[0000-0003-2757-535X]{Koki Inaba}
\affiliation{Department of Astronomy, Kyoto University, Kitashirakawa-Oiwake-cho, Sakyo-ku, Kyoto 606-8502, Japan}

\author[0000-0001-7821-6715]{Yoshihiro Ueda}
\affiliation{Department of Astronomy, Kyoto University, Kitashirakawa-Oiwake-cho, Sakyo-ku, Kyoto 606-8502, Japan}

\author[0000-0002-9754-3081]{Satoshi Yamada}
\affiliation{Institute of Physical and Chemical Research (RIKEN), 2-1 Hirosawa, Wako, Saitama 351-0198, Japan}
\affiliation{Department of Astronomy, Kyoto University, Kitashirakawa-Oiwake-cho, Sakyo-ku, Kyoto 606-8502, Japan}

\author[0000-0002-5701-0811]{Shoji Ogawa}
\affiliation{Department of Astronomy, Kyoto University, Kitashirakawa-Oiwake-cho, Sakyo-ku, Kyoto 606-8502, Japan}

\author[0000-0001-6653-779X]{Ryosuke Uematsu}
\affiliation{Department of Astronomy, Kyoto University, Kitashirakawa-Oiwake-cho, Sakyo-ku, Kyoto 606-8502, Japan}

\author[0000-0002-0114-5581]{Atsushi Tanimoto}
\affiliation{Graduate School of Science and Engineering, Kagoshima University, Kagoshima 890-0065, Japan}

\author[0000-0001-5231-2645]{Claudio Ricci}
\affiliation{N\'ucleo de Astronom\'ia de la Facultad de Ingenier\'ia, Universidad Diego Portales, Av. Ej\'ercito Libertador 441, Santiago, Chile}
\affiliation{Kavli Institute for Astronomy and Astrophysics, Peking University, Beijing 100871, People's Republic of China}





\begin{abstract}

We present the result of a broadband (0.5--70 keV) X-ray spectral
analysis of the late-merger galaxy Mrk~739, which contains a dual active galactic nucleus (AGN), Mrk~739E and Mrk~739W, with a separation of
~3.4 kpc.
The spectra obtained with NuSTAR, Chandra, XMM-Newton and
Swift/BAT are simultaneously analyzed
by separating the contributions from the two AGNs and extended
emission with the Chandra data.
To evaluate the reflection components from the AGN tori, 
we consider two models, a phenomenological one (\textsf{pexrav} and \textsf{zgauss}) and a more physically motivated one (XCLUMPY; \citealt{Tanimoto_2019}).
On the basis of the results with XCLUMPY,
we find that the AGNs in Mrk~739E and Mrk~739W have
intrinsic 2--10 keV luminosities of $1.0 \times 10^{43}$ and $7.5
\times 10^{41}\ \rm{erg}\ \rm{s}^{-1}$ absorbed by hydrogen column
densities of $N_{\rm{H}} < 6.5 \times 10^{19}\ \rm{cm}^{-2}$ and $N_{\rm{H}} = 6.9^{+3.2}_{-1.7} \times 10^{21}\ \rm{cm}^{-2}$, respectively. The torus
covering fraction of the material with $N_{\rm{H}} > 10^{22}
\rm{cm}^{-2}$ in Mrk~739E, $C_{\rm{T}}^{(22)} < 0.50$
at a 90\% confidence limit,
is found to be smaller than
those found for late-merger ultra/luminous infrared galaxies,  $C_{\rm{T}}^{(22)} = 0.71\pm0.16$ (mean and standard
deviation; \citealt{Yamada_2021}).
Considering the small star formation rate of Mrk~739E, we
suggest that the gas-to-mass ratio of the host galaxy is an important
parameter to determine the circumnuclear environment of an AGN in late
merger.

\end{abstract}

\keywords{Active galactic nuclei (16); Astrophysical black holes (98); High energy astrophysics (739); Supermassive black holes (1663); X-ray active galactic nuclei (2035)}


\section{Introduction} \label{sec:intro}

Past research since late 1990s has revealed 
that the masses
of supermassive black holes (SMBHs) in galactic centers
and those of galactic bulges are tightly 
correlated
(e.g., \citealt{Magorrian_1998}; \citealt{Ferrarese_2000}; \citealt{Kormendy_2013}).
These
results imply that
SMBHs and host galaxies have ``coevolved'' by regulating
their respective growths.
Galaxy mergers are thought to be one of the key paths
to explain this coevolution, because they can
facilitate both intense star
formation in the nuclei and efficient mass accretion onto the SMBHs
(e.g., \citealt{Hopkins_2008}).
Physical processes in the nuclear regions of merging galaxies
are however poorly understood.
Hence, 
investigating the circumnuclear environments
of active galactic nuclei (AGNs) in galaxy mergers is crucial
to elucidate the origin of the SMBH-galaxy coevolution.

Broadband X-ray observations are a powerful tool to study the
structure of AGNs because X-rays are tracers of all material
including gas and dust with various physical conditions.
A key component to understand the AGN feeding
mechanism is the so-called ``torus'' (e.g., \citealt{Urry_1995}; \citealt{Ramos_2017}),
which is considered as the fuel 
reservoir of the SMBH.
The spectral signatures of the X-ray radiation absorbed and/or reprocessed by
the torus
can provide constraints on its covering fraction and column density (e.g.,
\citealt{Kawamuro_2016};
\citealt{Tanimoto_2020};
\citealt{Toba2020};
\citealt{Uematsu2021};
\citealt{Tanimoto2022}).

Recently, \citet{Yamada_2021} have conducted a systematic
X-ray study of AGNs in 
57 local ultra/luminous infrared galaxy (U/LIRG) systems\footnote{Those with infrared (8--1000 $\mu$m) luminosities of
$L_{\rm IR} \geq 10^{12}L_{\odot}$ and of $L_{\rm IR} =
10^{11}$--$10^{12} L_{\odot}$ are called ULIRGs and LIRGs,
respectively \citep{Sanders_1996}.},
most of which are identified to be merging galaxies.
They find that the AGNs in merging U/LIRG systems become more deeply
``buried'' (i.e., covered by matter with a large solid angle)
with merging stage, confirming the implication by \citet{Ricci_2017a,Ricci_2021}
based on the absorption column-density distribution of local mergers (see also \citealt{Imanishi_2006,Imanishi_2008}; \citealt{Yamada_2019}).
In particular, it is found that AGNs in late-stage mergers 
with high Eddington ratios ($\log \lambda_{\rm Edd} > -1.5$)
show larger torus covering fractions than
those in nonmergers. These results
are in line with the theoretical
prediction that a galaxy merger triggers rapid mass accretion deeply
embedded by gas and dust (e.g., \citealt{Hopkins_2006}; \citealt{Blecha_2018}; \citealt{Kawaguchi_2020}).
However, since their sample is limited to U/LIRG populations, which have
high star formation rates (SFRs), the relation of the
``buried AGN'' nature with the host properties (e.g.,
star-formation rate, gas mass, stellar mass) remains unclear.

To test if galaxy mergers commonly produce
AGNs with large covering fractions 
regardless of their
SFRs, here
we focus on the AGNs in a late-stage merging galaxy that is {\it less
luminous} in the infrared band than U/LIRGs.
We choose Mrk~739 ($z=0.0297$) as our target. It is a merging
system consisting of Mrk~739E and Mrk~739W with 
a projected separation of $\approx$3.4 kpc, the
second smallest value among the Swift/BAT
dual-AGN sample by \citet{Koss_2012}. 
This defines this system
as a late-stage (stage C) merger according to the criteria of 
\citet{Stierwalt_2013}. However, unlike many other late-stage mergers,
Mrk~739 is not infrared luminous ($L_{\rm IR} \sim
10^{10.9}L_{\odot}$; e.g., \citealt{Imanishi_2014}) enough to be categorized as a LIRG, and hence
is not included in the sample of \citet{Yamada_2021}.
The optical spectrum of Mrk~739E shows broad emission lines, 
identifying it as a type-1 AGN (\citealt{Netzer_1987}, \citealt{Koss_2011a}, \citealt{Tubin_2021}).
\citet{Koss_2011a} suggest that Mrk~739W also contains an AGN
on the basis of 
the hard photon index in the soft X-ray (0.5--10 keV) band.
The optical spectra of Mrk~739W
show no evidence for broad emission lines, which may be 
contaminated by strong starburst components \citep{Tubin_2021}.

In this work, we aim to answer the question whether or not the AGN tori
in the Mrk~739 system have large covering fractions.
For this purpose, we conduct a broadband (0.5--70 keV) X-ray spectral
analysis of Mrk~739, utilizing the currently best available dataset
obtained with NuSTAR, Chandra, XMM-Newton, and Swift/BAT.
The NuSTAR data of Mrk~739 are presented here for the first time.

The paper is organized as follows. In Section~\ref{sec:obs_data},
we describe the observations and reduction of the data obtained with
the four satellites.
Section~\ref{sec:x-ray} describes the X-ray spectral model 
and give the results of spectral fitting.
To constrain the torus structure by assuming realistic geometry, we apply 
an X-ray clumpy torus model (XCLUMPY; \citealt{Tanimoto_2019}), 
which was also adopted in \citet{Yamada_2021}.
Finally, we discuss the properties of the AGN of Mrk~739E, which
is $\sim$15 times brighter than that of Mrk~739W, 
mainly focusing on its torus structure
in Section~\ref{sec:discussion}.
Section~\ref{sec:conclusion} reports the conclusions of this work.
In this paper, we assume the solar abundances by
\citet{Anders_1989}. Throughout this paper, we adopt the cosmological parameters of $H_0 = 70\ \rm{km}\ \rm{s}^{-1}\ \rm{Mpc}^{-1}$, $\Omega_{\rm{M}} = 0.3$, and $\Omega_{\Lambda}=0.7$. Errors on spectral parameters correspond
to 90\% confidence limits for a single parameter of interest.

\section{OBSERVATIONS AND DATA REDUCTION} \label{sec:obs_data}

The observation log of the X-ray spectral data used in this paper
(NuSTAR, Chandra, XMM-Newton, and Swift/BAT) is summarized in
Table~\ref{tab1-obs}. We conducted data reduction
in accordance with the procedures described below.

\subsection{NuSTAR} \label{subsec:NuSTAR}

NuSTAR \citep{Harrison_2013} observed Mrk~739 with the two focal plane
modules (FPMs: FPMA and FPMB), which are sensitive to X-ray energies
in the 3--79~keV band. The observation was performed on 2017 March 16 for an
exposure time of 18.5 ks.  The data presented here were reduced by
using HEAsoft v6.28 and CALDB v20201217.  FPMs' data were reprocessed
by using \textsc{nupipeline} and \textsc{nuproducts}.

We extracted spectral data from a circular region with a radius
of 60\arcsec, and subtracted the background taken from an annulus with
inner and outer radii of 60\arcsec\ and 120\arcsec\ surrounding the source
concentrically.  
Considering the limited photon statistics,
we coadded the source spectra, background spectra, RMF, and ARF of the two FPMs by using \textsc{addascaspec}.
We grouped bins so that each of them included at
least 50 photon counts.
The angular resolution of NuSTAR is 58\arcsec\ in
full width at half maximum (FWHM), which is not sufficient 
to separate
the AGNs in the two galaxies (Mrk~739E and Mrk~739W) of the system. 
Hence, our NuSTAR spectrum contains emission from both AGNs and galaxies.

\subsection{Chandra} \label{subsec:Chandra}
Chandra \citep{Weisskopf_2002} observed Mrk~739 with the Advanced CCD
Imaging Spectrometer (ACIS: \citealt{Garmire_2003}).
The observation was performed
on 2011 April 22 for an exposure time of 13.0 ks. 
The data were reduced in the standard manner using the Chandra Interactive Analysis of Observations (CIAO v4.12) software and calibration files (CALDB v4.9.2.1).
Thanks to its good angular resolution ($<$1\arcsec),
Chandra was able to spatially resolve the two galaxies and the extended
soft X-ray emission surrounding them \citep{Koss_2011a}.

\begin{figure*}
\epsscale{0.90}
\plottwo{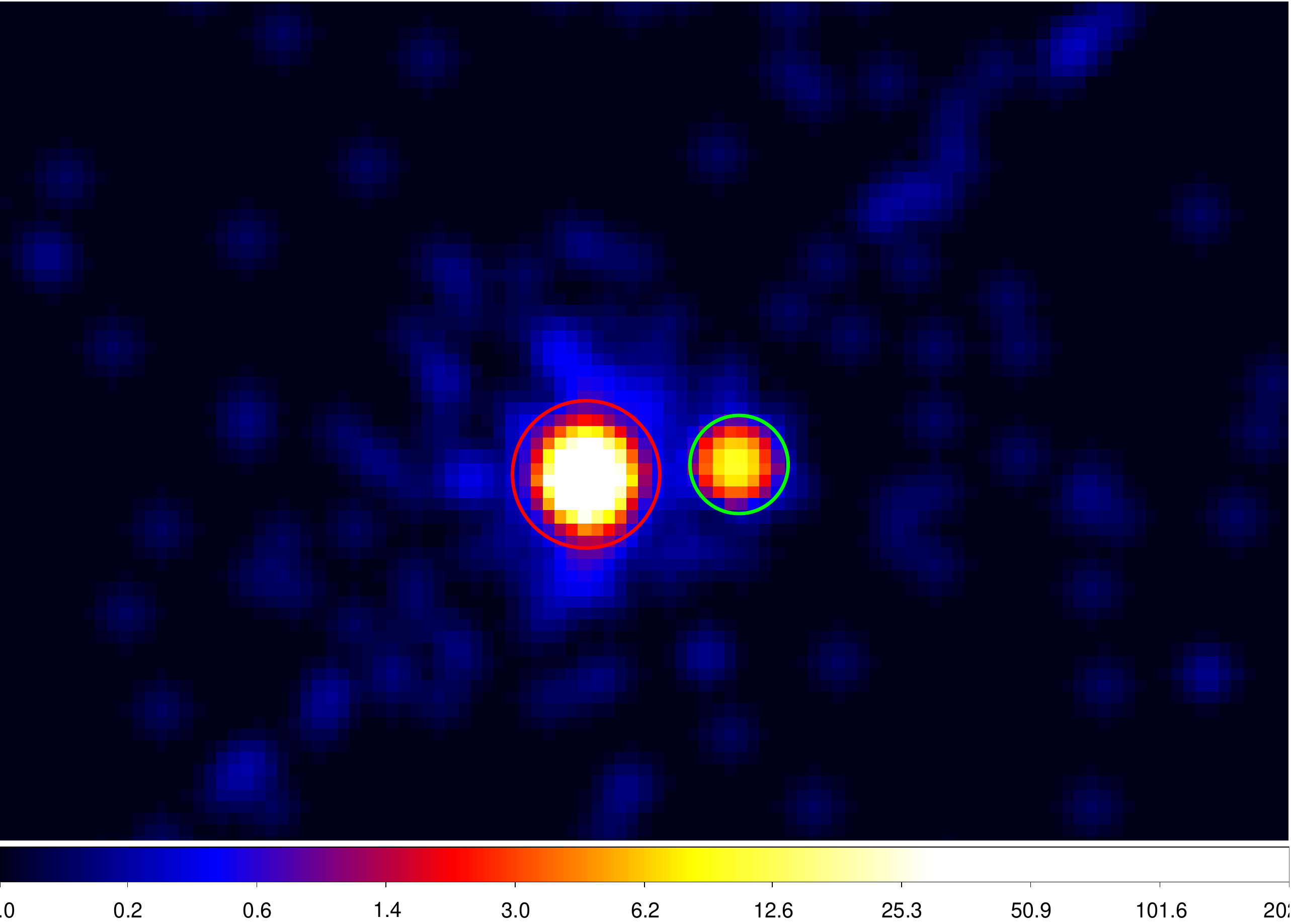}{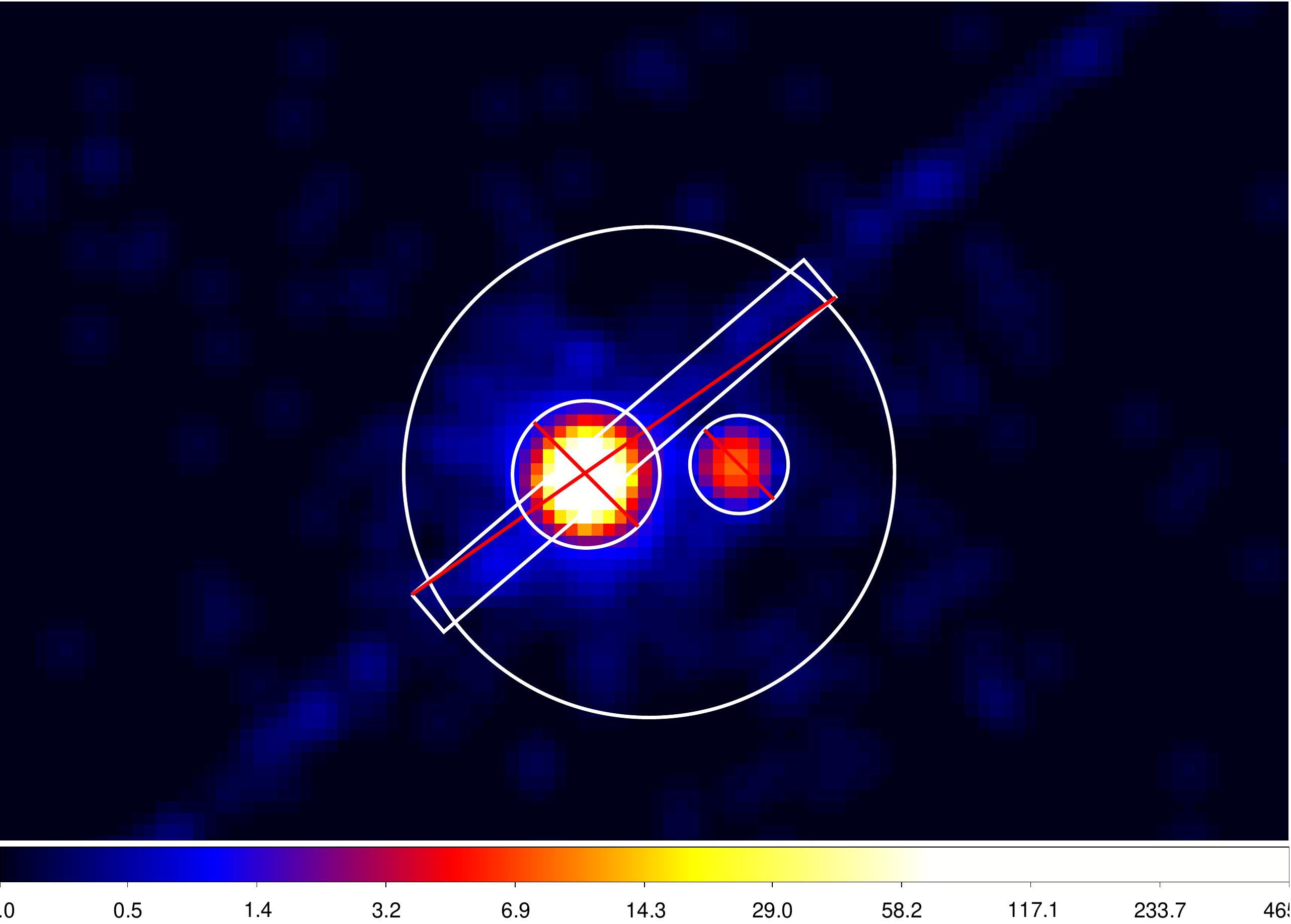}
\caption{(Left) Chandra/ACIS-S image 
of the central region of the
Mrk~739 system in the 2.0-8.0 keV band, smoothed with a 2D Gaussian of a $\sigma$ radius of 1.5 pixels (each pixel is 0.492\arcsec).
The spectral extraction regions for Mrk~739E and Mrk~739W
are marked by the left (red) and right (green) circles,
respectively.
(Right) The same but in the 0.4-2.0 keV band.
  The large circle (white) marks the
the spectral region for the extended emission from which the rectangle region containing a readout streak 
and the two circular regions in the left panel are excluded.
} \label{fig-ds9}
\end{figure*}

Thus,
we extracted three spectral regions: two circular regions with radii of 3\arcsec\ and 2\arcsec, respectively, centered on the galactic centers of Mrk~739E and Mrk~739W. 
For the extended emission, we adopt a circular region with a radius of 10\arcsec\ 
from which the two circular regions mentioned above were excluded (see Figure~\ref{fig-ds9}). 
Additionally,
we excluded a rectangle region (2\arcsec$\times$10\arcsec) where instrumental
noises are prominent.
We subtracted the background, which was taken from a source-free circular region with a radius of 10\arcsec\ on the same chip.
We grouped bins of the spectra of
Mrk~739E, Mrk~739W, and the soft extended emission, respectively, so that each bin included at least 20 photon counts.

\subsection{XMM-Newton} \label{subsec:XMM-Newton}
XMM-Newton \citep{Jansen_2001} observed Mrk~739
on 2009 June 14 for an exposure time of 11.9 ks
with the European Photon Imaging Cameras (EPICs).
We reduced the data with the science analysis software (SAS) v18.0.0 and current calibration files (CCF) released on 2020 May 14.
In this paper, we only utilized the data of 
pn \citep{Struder_2001}, which has larger effective area
than MOS1 and MOS2. 

The angular resolution of XMM-Newton, 15\arcsec\ (FWHM), is not
small enough to separate the two galactic centers of Mrk~739.
Thus, we extracted spectral data from a single
circular region with a radius of 40\arcsec.
The background was taken from a circular source-free region with a radius of 40\arcsec\ on the same chip.
We grouped bins of source so that at least 50 photon counts
were included per bin.

\subsection{Swift/BAT} \label{subsec:Swift/BAT}

We also utilized the hard X-ray spectrum of Mrk~739 obtained with
Burst Alert Telescope (BAT) on
the Neil Gehrels Swift observatory \citep{Gehrels_2004}
averaged over 105 months \citep{Oh_2018} from 2004 December to 2013 August.

\begin{deluxetable*}{ccrrc}
\tablewidth{\textwidth}

\tablecaption{Summary of Observations \label{tab1-obs}}
\tablehead{
\colhead{Satellite}      &
\colhead{ObsID}          &
\colhead{Start Date (UT)} &
\colhead{End Date (UT)}   &
\colhead{Net exposure (ks)\tablenotemark{a}} 
}
\startdata
NuSTAR & 60260008002 & 2017-03-16 03:56:09 & 2017-03-16 09:05:16 & 18.5 \\
Chandra (ASIS-S) & 12863 & 2011-04-22 02:26:55 & 2011-04-22 06:03:45 & 13.0 \\
XMM-Newton & 0601780401 & 2009-06-14 08:23:45 & 2009-06-14 11:42:21 & 11.9 \\
Swift/BAT & - & 2004 December & 2013 August & -
\enddata
\tablenotetext{a}{Based on the good time interval of FPMA for NuSTAR and EPIC-pn for XMM-Newton.}

\end{deluxetable*}

\section{X-RAY SPECTRAL ANALYSIS} \label{sec:x-ray}

To best constrain the properties of the AGNs,
we take the same approach as adopted by \citet{Yamada_2018} for the
local merging galaxy Mrk~463.
We perform simultaneous fit to the six spectra:
the individual spectra of Mrk~739E, Mrk~739W,
and the extended emission observed with Chandra/ACIS-S (in the 0.5--7.0 keV,
0.8--5.0 keV, 0.5--3.0 keV bands, respectively), and the total spectra
from the whole Mrk~739 system observed with NuSTAR/FPMs
(3.0--60 keV), XMM-Newton/pn (0.5--10 keV), and Swift/BAT (20--70 keV). 
Here we adopt the energy intervals where the spectra
have sufficiently high signal-to-noise ratios.
Spectral fitting is performed on
XSPEC \citep{Arnaud_1996} v12.11.1
based on $\chi^2$-statistics. We always consider the Galactic absorption
(total of \ion{H}{1} and H$_2$),  whose hydrogen column density is fixed
at $2.14\times 10^{20}\ \rm{cm}^{-2}$ \citep{Willingale_2013}.

\subsection{AGN Spectral Model} \label{subsec:model}

In this section, we
describe the details of spectral models for the AGNs in Mrk
739E and Mrk~739W.
We first begin with a simple phenomenological model (Model~I). 
Then, we apply a more physically motivated one (Model~II) to constrain
the torus parameters.

\subsubsection{Model~I}

A typical X-ray spectrum of an AGN (e.g., \citealt{Ogawa_2019})
consists of a direct component, which is often subject to absorption
by line of sight matter (e.g., warm absorbers, torus), 
reflected components from the torus and/or the accretion disk
accompanied by fluorescence lines,
and a Thomson-scattered component (plus emission from a photoionized
plasma).
In addition to the AGN emission, optically-thin thermal emission
from the host galaxy is often present.

Thus, we consider these AGN components for the Mrk 739E and Mrk
739W spectra except for Thomson-scattered ones, which are not 
significantly required in our data.
Firstly, we perform spectral fitting without any reflection components.
We find, however, that the model systematically underproduces
the hard X-ray flux above 10~keV by a factor of 2, 
suggesting the presence of a reflection hump 
by the torus and/or the accretion disk in Mrk~739.
We also see a hint of a narrow emission line around 6.4~keV
in the XMM-Newton spectrum.
Accordingly, 
we add the \textsf{pexrav} \citep{Magdziarz1995} and \textsf{zgauss} components to represent the reflection continua from 
cold matter and Fe K$\alpha$ fluorescence lines, respectively.
The models for Mrk~739E and Mrk~739W are expressed as follows in XSPEC terminology:
\begin{eqnarray}
\label{equation:mrk739e_1}
\mathrm{Mrk\ 739E} &=& \textsf{const0*phabs} \nonumber\\
&*& (\textsf{const1*zphabs*cabs} \nonumber\\
&*& \textsf{mtable\{xout\_mtable.fits\}} \nonumber\\
&*& \textsf{zcutoffpl} \nonumber\\
&+& \textsf{pexrav + zgauss} \nonumber\\
&+& \textsf{apec})
\end{eqnarray}
\begin{eqnarray}
\label{equation:mrk739w}
\mathrm{Mrk\ 739W} &=& \textsf{const0*phabs} \nonumber\\
&*& (\textsf{const1*zphabs*cabs} \nonumber\\
&*& \textsf{zcutoffpl} \nonumber\\
&+& \textsf{pexrav + zgauss})
\end{eqnarray}
\begin{enumerate}
\renewcommand{\labelenumi}{(\arabic{enumi})}
\item The \textsf{const0} and \textsf{phabs} terms represent the cross calibration factor of among different instruments and the Galactic absorption, respectively. 
We set \textsf{const0} of NuSTAR to unity as the reference.
  We also fix that of Swift/BAT at unity, which 
would be strongly coupled with \textsf{const1} otherwise.
Those of Chandra and XMM-Newton are set to the values derived by  \citet{Madsen_2017}, 1.09 and 0.89, considering the limited photon statistics.
\item The \textsf{const1} factor takes into account time variability of Mrk~739E among different observational epochs.
We ignore time variability of Mrk~739W (i.e., \textsf{const1} is set to unity) because the flux of Mrk~739W is too faint to detect any variability. 
To confirm the validity of this assumption, we 
also perform spectral fitting by adopting the \textsf{const1} values of 0.3 or 3.0 (a typical variability range of nearby AGNs over $\sim$10 years; \citealt{Kawamuro_2016}) 
for Chandra, XMM-Newton, and Swift/BAT data.
Except for the two cases where the $\chi^2$ values are unacceptably increased by $\gtrsim$30, 
we find that the choices of \textsf{const1} values do not affect our main conclusion.
The \textsf{zphabs} and \textsf{cabs} terms represent
photoelectric absorption and Compton scattering,
respectively. The hydrogen column density along the
line of sight is fixed at
$N_{\rm{H}}^{\rm{LOS}}=10^{19}\ \rm{cm}^{-2}$
(the best-fit value
obtained with Model~II, see Section~3.3.2.),
which cannot be well
constrained with Model~I. 
\item The multiplicative table model
\\
\textsf{mtable\{xout\_mtable.fits\}} represents absorptions by ionized matter, often called ``warm absorbers''. To generate the table model, we utilized XSTAR v2.54a (\citealt{Kallman_Bautista_2001}; \citealt{Bautista_Kallman_2001}) assuming the solar abundances.
We assume that the ionization parameter ($\xi$) and hydrogen column density ($N_{\rm H}$) are variable among different epochs. Since NuSTAR does not
cover the soft X-ray band, we adopt the same parameters as determined by the Chandra data for the NuSTAR spectrum.
\item The \textsf{zcutoffpl} term is a direct component (a power-law with an exponential high energy cutoff). Here, we assume that the power-law photon index did not vary during the different observational epochs. The cutoff energy is fixed at 160 keV as a typical value of AGNs \citep{Ricci_2018}.
The \textsf{pexrav} term represents a reflection component from optically-thick cold matter \citep{Magdziarz1995}. 
\textsf{pexrav} has eight parameters, the power law photon index ($\Gamma$), the cutoff energy ($E_{\rm{cut}}$), the reflection strength ($R\equiv \Omega/2\pi$, where $\Omega$ is the solid angle of the reflector), the redshift ($z$), the abundance ($Z$; elements heavier than He), the iron abundance $(Z_{\rm Fe})$, the inclination angle ($i$), and the normalization ($K_{\rm{ref}}$).
We link $\Gamma$, $E_{\rm{cut}}$, and $K_{\rm{ref}}$ to those of the direct component.
We assume that $Z=1$ and $Z_{\rm{Fe}}=1$ (i.e., solar abundances).
The inclinations of Mrk~739E and Mrk~739W are fixed at $i=45^{\circ}$ and $i=60^{\circ}$, respectively 
(see the description of Model~II component (5) in Equation~(\ref{equation:mrk739e})).
We set $R$ of Mrk~739E to be free within a negative range ($-2\leq R \leq 0$) to obtain only the reflected component.
Due to the limited photon statistics, we fix $R$ of Mrk~739W at $-0.8$, the average value for nearby moderately-obscured AGNs \citep{Kawamuro_2016}.
The \textsf{zgauss} term represents an Fe K$\alpha$ fluorescence line,
whose center energy and 1$\sigma$ width are fixed at 6.4~keV and 
of 20~eV, respectively.
While we set the normalization of Mrk~739E to be free, we fix that of Mrk~739W so that the Fe K$\alpha$ lines of the two AGNs have similar equivalent widths
with respect to the \textsf{pexrav} components.
\item The last term, \textsf{apec}, represents optically-thin thermal plasma
due to the star formation activity in the host galaxy.
We assume the parameters to be invariable among different epochs.
\end{enumerate}
In the spectral model of Mrk~739W, we do not include the warm-absorber
and optically-thin thermal plasma, which are
not significantly required from our spectra.
The analysis results with Model~I are mentioned in Section~\ref{subsec:result}.

\subsubsection{Model~II}

As a more physically motivated model, we replace the torus reflection component \textsf{pexrav} and \textsf{zgauss} in Model~I
with the X-ray clumpy torus model XCLUMPY \citep{Tanimoto_2019}. This enables us to constrain the torus structure and to compare our results with those of AGNs in late mergers obtained by \citet{Yamada_2021}.

We find, however, that a hard X-ray excess around 20~keV
remains with this model.
This is because 
XCLUMPY whose parameters are mainly constrained by the line-of-sight
absorption and the narrow Fe K$\alpha$ equivalent width 
produces a much weaker hard X-ray
reflection hump than the \textsf{pexrav} component in Model~I.
This suggests the presence of a relativistically blurred reflection
component from the accretion disk that produces a strong reflection
hump at $\sim$20 keV but not a ``narrow'' Fe K$\alpha$ line (see also Section~\ref{subsec:result}).
Accordingly, we added a \textsf{relxill} component for Mrk~739E. 
The final spectral models of Mrk~739E and Mrk~739W are
expressed as follows in XSPEC terminology: 
\begin{eqnarray}
\label{equation:mrk739e}
\mathrm{Mrk\ 739E} &=& \textsf{const0*phabs} \nonumber\\
&*& (\textsf{const1*zphabs*cabs} \nonumber\\
&*& \textsf{mtable\{xout\_mtable.fits\}} \nonumber\\
&*& \textsf{(zcutoffpl + relxill)} \nonumber\\
&+& \textsf{atable\{xclumpy\_v01\_RC.fits\}} \nonumber\\
&+& \textsf{atable\{xclumpy\_v01\_RL.fits\}} \nonumber\\
&+& \textsf{apec})
\end{eqnarray}

\begin{eqnarray}
\label{equation:mrk739w}
\mathrm{Mrk\ 739W} &=& \textsf{const0*phabs} \nonumber\\
&*& (\textsf{const1*zphabs*cabs} \nonumber\\
&*& \textsf{zcutoffpl} \nonumber\\
&+& \textsf{atable\{xclumpy\_v01\_RC.fits\}} \nonumber\\
&+& \textsf{atable\{xclumpy\_v01\_RL.fits\}})
\end{eqnarray}

\begin{enumerate}
\renewcommand{\labelenumi}{(\arabic{enumi})}
\item Same as the component (1) of Equation~(\ref{equation:mrk739e_1}).

\item Same as the component (2) of Equation~(\ref{equation:mrk739e_1}).
Here, the hydrogen column density along the line of sight
is not a free parameter, but determined by the torus parameter (see Equation~\ref{N_H}).

\item Same as the component (3) of Equation~(\ref{equation:mrk739e_1}).

\item Same as the component (4) of Equation~(\ref{equation:mrk739e_1}) in terms of \textsf{zcutoffpl}. 
The \textsf{relxill} term represents a relativistic reflection component from the accretion disk \citep{Dauser_2013,Garc_a_2014}. This model (RELXILL) combines
the XILLVER code \citep{Garc_a_2014}, which calculates a reflected spectrum from the accretion disk, and the RELLINE code \citep{Dauser_2010,Dauser_2013}, which takes into account relativistic smearing effects.
RELXILL has 14 parameters: 
the inner $(R_{\rm in})$ and outer $(R_{\rm out})$ radii of the disk,
the boundary radius $(R_{\rm br})$ at which the emissivity index changes,
the index $(q_1)$ which gives emissivity as $r^{-q_1}$ between $R_{\rm in}$ and $R_{\rm br}$,
the index $(q_2)$ which gives emissivity as $r^{-q_2}$ between $R_{\rm br}$ and $R_{\rm out}$,
the spin parameter of the black hole $(a)$,
the inclination angle $(i)$ between the line of sight and the line perpendicular to the disk,
the redshift $(z)$,
the power-law index $(\Gamma)$,
the ionization parameter of the disk $(\xi)$,
the iron abundance $(Z_{\rm Fe})$,
the cutoff energy $(E_{\rm cut})$,
the ratio $(R)$ of the solid angle of the reflector to $2\pi$\footnote{When this parameter is set to a negative value, only the reflected component is returned},
and the normalization $(\rm{K_{\rm rel}})$.
The photon index ($\Gamma$) and cutoff energy ($E_{\rm cut}$) are linked between the direct component and RELXILL.
The inclination angle ($i$) is fixed at the same value as in the \textsf{pexrav} model (see the component (5)).
We assume that $R_{\rm in} = R_{\rm ISCO}$,
$R_{\rm out} = 10^3 \, r_{\rm g}$ ($r_{\rm g}$ is the gravitational radius), 
$q_1 = q_2 = 2.4$, which is a mean value for local Seyfert~1 galaxies \citep{Patrick_2012}, 
$a=0$, $\xi=0$, and $Z_{\rm Fe} = 1.0 $ (i.e., solar abundances),
because these parameters cannot be well constrained by our data.
We set $R=-0.25$ to obtain only the reflected component.
Thus, the free parameter of RELXILL in our fitting is only
the normalization $(\rm{K_{\rm rel}})$.
Since both \textsf{zcutoffpl} and \textsf{relxill} components are emitted
from a compact region much smaller than light years,
we assume that they are subject to the same
absorption and time variability factor (\textsf{const1}).

\item The \textsf{atable\{xclumpy\_v01\_RC.fits\}} and \textsf{atable\{xclumpy\_v01\_RL.fits\}} term correspond to the
reflection continuum and fluorescence lines
from the torus, respectively, based on the
XCLUMPY model \citep{Tanimoto_2019}.
The parameters of XCLUMPY are the hydrogen column density along the equatorial plane $(N^{\rm Equ}_{\rm H})$, the torus angular width $(\sigma)$,
the inclination angle $(i)$, 
the power-law photon index $(\Gamma)$, the cutoff energy $(E_{\rm cut})$, the redshift $(z)$, and the normalization $(\rm{K_{\rm ref}})$.
We link $\Gamma$, $E_{\rm cut}$,
and $\rm{K_{\rm ref}}$ to those of the direct component.
The inclination of Mrk~739E is fixed at $i=45^{\circ}$,
a typical value for Seyfert~1 galaxies as constrained by the XCLUMPY model \citep{Ogawa_2021}.
The inclination of Mrk~739W is fixed at $i=60^{\circ}$
because a lower $i$ value makes $N^{\rm Equ}_{\rm H}$ pegged at its upper boundary ($10^{25}\ \rm{cm}^{-2}$).
Thus, the free parameters are $N^{\rm Equ}_{\rm H}$ and $\sigma$.
Because of coupling between the two parameters, however,
we find it difficult to determine both values simultaneously. Hence,
we conduct three patterns of spectral fitting
by fixing $\sigma$ of Mrk~739E at $10^{\circ}$,
$15^{\circ}$, or $20^{\circ}$\footnote{
We note that $\sigma=10^\circ$ corresponds to the lower boundary
of the parameter range in XCLUMPY, and adopting $\sigma\geq 25^\circ$ makes
$N^{\rm Equ}_{\rm H}$ pegged at its lower boundary ($10^{23}$ cm$^{-2}$).
}.
The torus angular width $\sigma$ of Mrk~739W is fixed at $15^{\circ}$, which cannot be well constrained.
From $N^{\rm Equ}_{\rm H}$ and
$\sigma$, the hydrogen column density along the line of sight ($N^{\rm LOS}_{\rm H}$) at the
elevation angle $\theta$ ($=90^{\circ}-i$) is calculated as follows:
\begin{eqnarray} \label{N_H}
N^{\rm LOS}_{\rm H}\left(\theta\right) = N^{\rm Equ}_{\rm H} \exp\left(-\left(\frac{\theta}{\sigma}\right)^2\right).
\end{eqnarray}
This value of $N^{\rm LOS}_{\rm H}(\theta)$ is applied to the absorption terms
in (2). 
Considering that the large size of the torus ($\sim$pc), 
we assume that the flux of the torus reflection component
is invariable among the different observation epochs,
which were separated by 2--8 years.

\item Same as the component (5) of Equation~(\ref{equation:mrk739e_1}).
\end{enumerate}

The analysis results with Model~II are mentioned in Section~3.3.2.

\subsection{Extended Emission}
Generally, extended emission around an AGN consists of scattered/photoionized  component from the AGN 
and optically-thin thermal emission from the host galaxy.
Since we find that thermal emission modelled by \textsf{apec} is not required by our data, we employ only a simple power-law as the scattered AGN component. 
The final model of the extended emission is expressed as follows in XSPEC
terminology:
\begin{eqnarray}
\label{equation:extended}
\mathrm{Extended} &=& \textsf{const0*phabs} \nonumber\\
&*& \textsf{zcutoffpl}
\end{eqnarray}
\begin{enumerate}
\renewcommand{\labelenumi}{(\arabic{enumi})}
\item Same as Mrk~739E component (1).
\item 
The \textsf{zcutoffpl} term represents scattered components from 
the AGNs in Mrk~739E and Mrk~739W.  This
component is assumed to be constant over all the observations. We
correct for the effect of excluding the rectangle area in the spectral
extraction region of the Chandra spectrum by assuming a constant
surface brightness.

\end{enumerate}

\subsection{Results} \label{subsec:result}
\subsubsection{Results-of-Model-I}

The phenomenological model, Model~I, is able to reproduce the
observed spectra of the six instruments simultaneously. 
Table~\ref{relxill-parameters} (Model I) summarizes the best-fit parameters.
In Mrk 739E, time-variable warm absorbers with
$\log N_{\mathrm{H}}/\mathrm{cm}^2 \sim$ (21.5--22.0) are detected.
Line-of-sight absorption by neutral material with 
$\log N_{\mathrm{H}}/\mathrm{cm}^2 =21.9^{+0.19}_{-0.17}$
is 
required in Mrk~739W.
An $F$-test verifies that
the addition of the \textsf{pexrav} and \textsf{zgauss} components 
significantly improves
the fit at a ${>}99\%$ confidence level
($\chi^2$/d.o.f = $649.0/618$ from $\chi^2$/d.o.f = $679.9/620$).
We obtain a reflection strength of $R= -1.12^{+0.48}_{-0.66}$ and
an equivalent width of a narrow Fe K$\alpha$ line of $55\pm46$~eV for
Mrk 739E. We note that this equivalent width is smaller than
the theoretical prediction from the observed \textsf{pexrav} component
($\gtrsim$110 eV) as estimated with the \textsf{pexmon} model \citep{Nandra_2007}.
This may support the presence of a relativistic reflection
component introduced in Model~II (Section~3.1.2). 

\subsubsection{Results-of-Model-II}

\begin{figure*}
\epsscale{1.17}
\plottwo{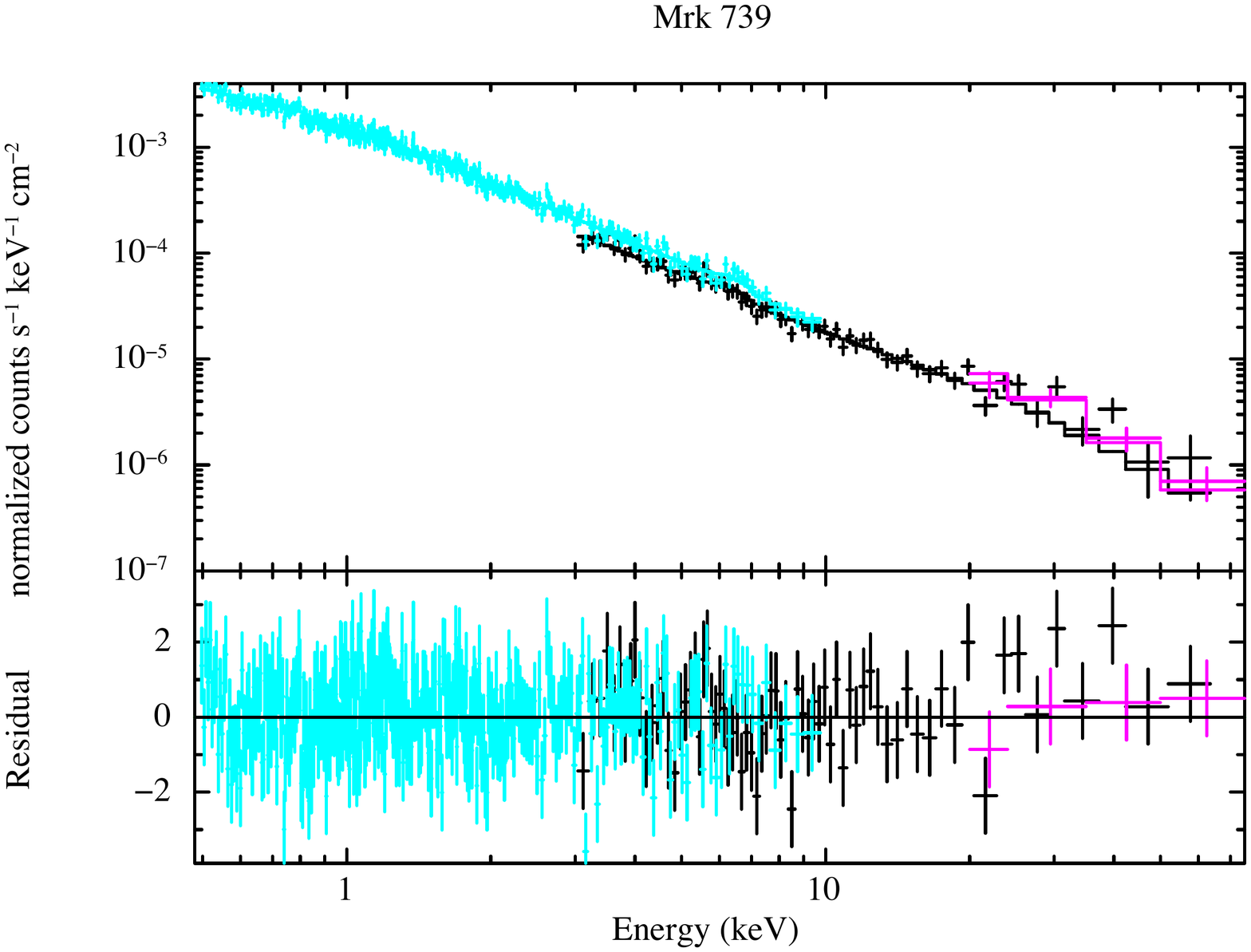}{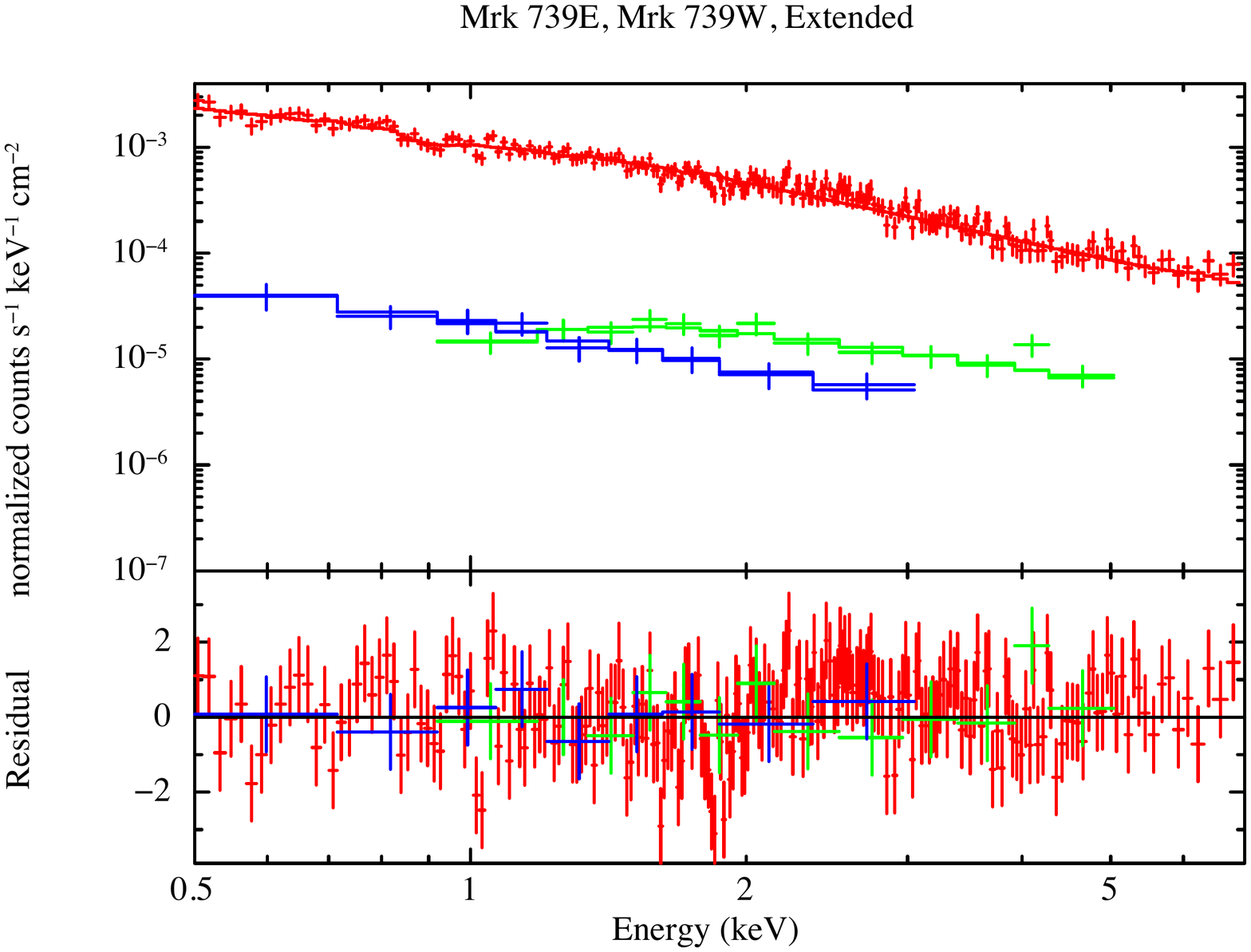}
\caption{
The observed spectra of Mrk~739
folded with the energy responses but corrected for the effective
area (crosses). 
The best-fit models are overplotted (lines).
(Left) The black, light blue, and magenta crosses and lines represent the NuSTAR, XMM-Newton, and Swift/BAT data of the total spectrum of Mrk~739, respectively.
(Right) The red, green, and blue crosses and lines represent the Chandra 
spectra of Mrk~739E, Mrk~739W, and the extended emission,
respectively. 
The residuals are plotted in the lower panels.
}
\label{fig-fitting}
\end{figure*}

\begin{figure*}
\epsscale{1.17}
\plottwo{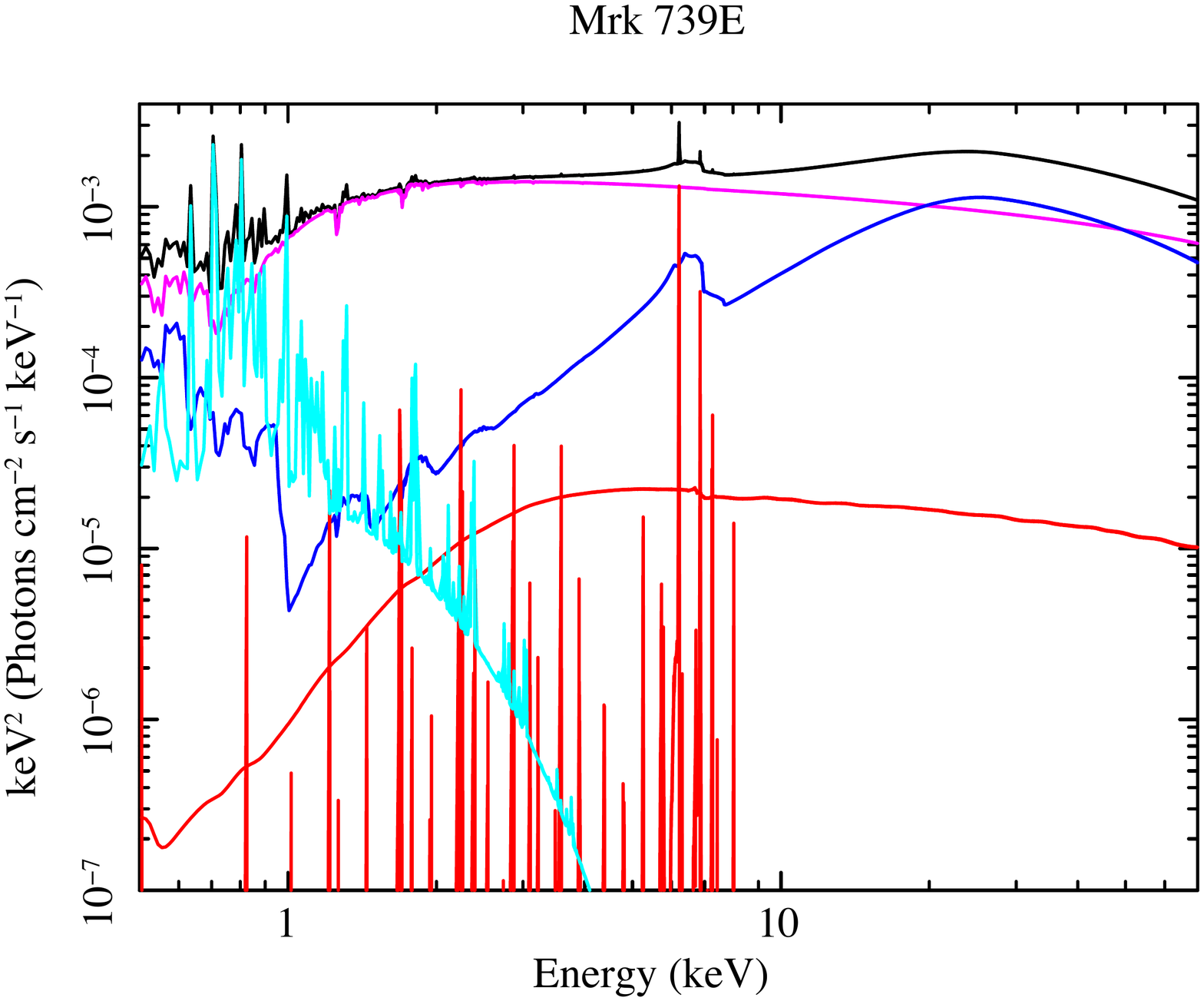}{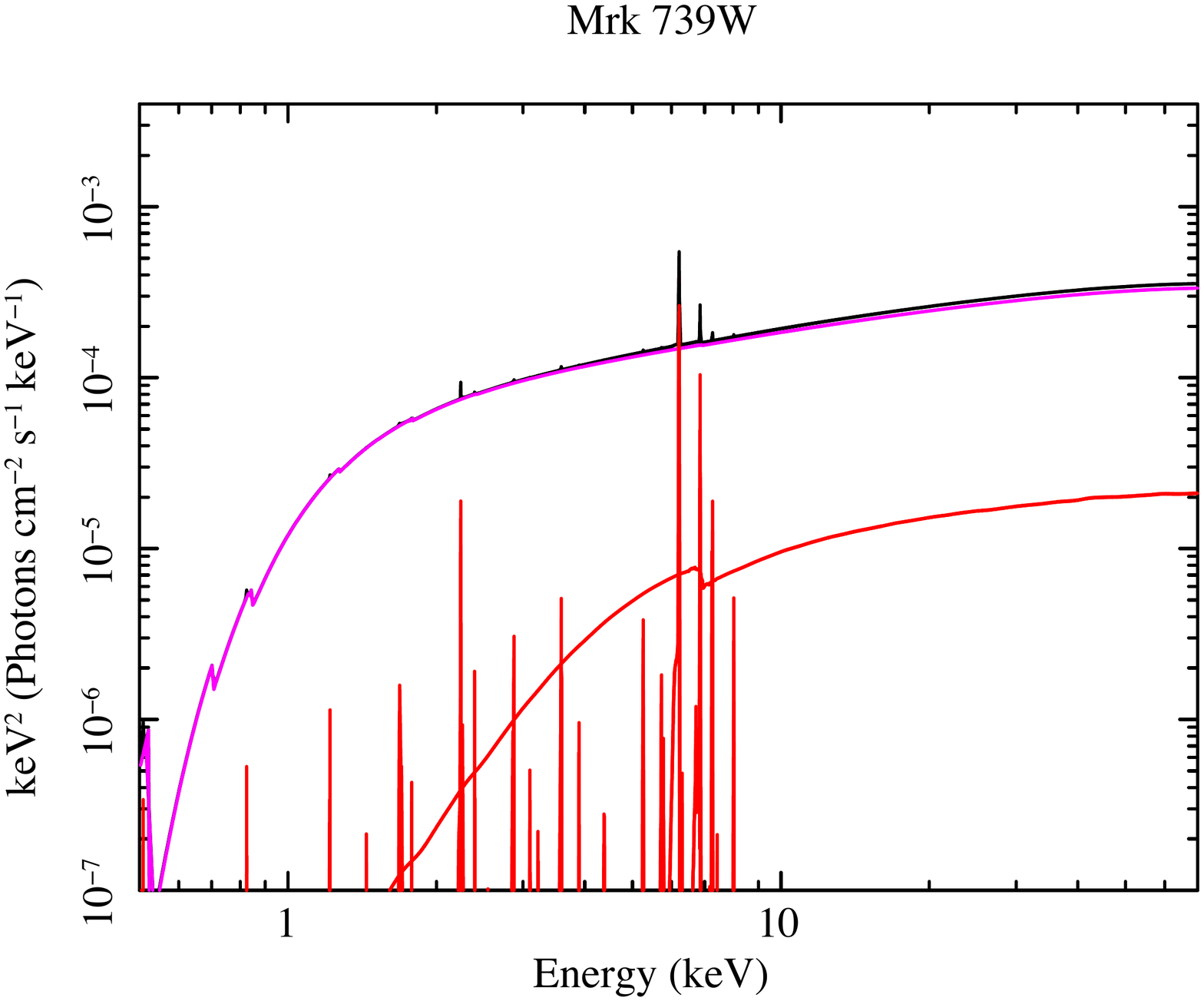}
\caption{
The best-fit models in units of $E I_E$, where $I_E$ is the energy flux at the energy $E$. The black, magenta, red, blue, and cyan lines represent the total model, direct component, reflection component from the torus, reflection component from the accretion disk, and optically-thin thermal emission, respectively. 
(Left) The best-fit model of Mrk~739E at the epoch of
the Chandra observation. 
(Right) The best-fit model of Mrk~739W. 
}
\label{fig-model}
\end{figure*}

We find that Model~II also well reproduces the data.
Table~\ref{relxill-parameters} (Model II)
summarizes the best-fit parameters for the case of $\sigma=15^\circ$.
The presence of the variable warm absorbers in Mrk 739E is confirmed.
An $F$-test confirms that the addition of the \textsf{relxill} term provides a significantly better fit at a ${>}99\%$ confidence level
($\chi^2$/d.o.f = $644.7/618$ from $\chi^2$/d.o.f = $665.4/619$).
The observed spectra and the best-fit model are
plotted in Figure~\ref{fig-fitting} and Figure~\ref{fig-model},
respectively.
As mentioned above, 
the main features in the observed spectra that constrain the torus
parameters of XCLUMPY are the line of sight column density (or its
upper limit) and the equivalent width of the Fe K$\alpha$ line. As
evident from Equation~(\ref{N_H}) and Equation~(\ref{C_T})
of this paper and Figure~3 of \citet{Ogawa_2019},
respectively, these constraints inevitably cause parameter coupling
(anti correlation) between $N^{\rm Equ}_{\rm H}$ and $\sigma$. To check if our main
results would be affected by this,
we also perform spectral fitting by
fixing $\sigma$ at $10^{\circ}$ or $20^{\circ}$ for Mrk~739E. We
confirm that the spectra are also well reproduced. The best-fit values
of $N^{\rm Equ}_{\rm H}$ and reduced $\chi^2$ values for the three cases
($\sigma=10^\circ, 15^\circ, 20^\circ$) are summarized in the
Table~\ref{sigma-NH-chi}.

\begin{deluxetable*}{lllcccc}
\tablewidth{\textwidth}
\tablecaption{Best-fit Spectral Parameters ($\sigma = 15$) \label{relxill-parameters}}

\tablehead{
\colhead{Region}           &
\colhead{Component}        &
\colhead{No.}              &
\colhead{Parameter}        &
\colhead{Model I}  &
\colhead{Model II}  &
\colhead{Units}            
}
\startdata
{Mrk~739E} &{\tt mtable\{xout\_mtable.fits\}}& (1)  &$N_{\rm{H}}$ &$8.57^{+1.02}_{-1.12},\ 3.06^{+0.68}_{-0.62}$&$8.80^{+1.04}_{-1.01},\ 3.79^{+0.81}_{-0.78}$    &$10^{21}$ cm$^{-2}$                      \\
      && (2)  &$\log \xi$  &$0.89^{+0.07}_{-0.09},\ 0.58^{+0.09}_{-0.10}$&$0.74^{+0.07}_{-0.06},\ 0.41^{+0.06}_{-0.17}$ &    \\
      &{\tt zcutoffpl}& (3) &$\Gamma_{\rm{AGN}}$                       &$2.07^{+0.06}_{-0.05}$&$2.15^{+0.09}_{-0.06}$                &      \\
      && (4)  &$K_{\rm{AGN}}$             &$1.81^{+0.16}_{-0.15}$&$1.91^{+0.19}_{-0.15}$
      &$10^{-3}$ keV$^{-1}$ cm$^{-2}$ s$^{-1}$  \\
      &{\tt relxill}& (5)  &$K_{\rm{RELXILL}}$            &\nodata&$5.25^{+2.92}_{-1.70}$    &$10^{-5}$ keV$^{-1}$ cm$^{-2}$ s$^{-1}$ \\
      &{\tt pexrav}& (6)  &$R$            &$-1.12^{+0.48}_{-0.66}$&\nodata    & \\
      &{\tt zgauss}& (7)  &$K_{\rm{Fe}}$            &$2.66^{+2.22}_{-2.22}$&\nodata    & $10^{-6}$ keV$^{-1}$ cm$^{-2}$ s$^{-1}$\\
      &{\tt atable\{xclumpy\_v01\_RC(L).fits\}}& (8)  &$N^{\rm{Equ}}_{\rm{H}}$       &\nodata&$<5.30$ &$10^{23}$ cm$^{-2}$     \\
      && (9)  &$\sigma$    &\nodata&$15.0\ (\rm{fixed})$ 
      &$\rm{degree}$                                      \\
      &{\tt apec}& (10)  &$K_{\rm{apec}}$            &$1.29^{+0.26}_{-0.21}$&$1.46^{+0.42}_{-0.27}$ &$10^{-4}$ cm$^{-5}$                \\
      && (11)  &$k$T                 &$0.51^{+0.06}_{-0.08}$&$0.45^{+0.07}_{-0.09}$          
      &keV                                 \\
      &{\tt const1}& (12) &$N_{\rm{Chandra}}$               &$1.26^{+0.08}_{-0.07}$&$1.25^{+0.07}_{-0.07}$         &                           \\
      && (13) &$N_{\rm{XMM}}$               &$1.34^{+0.08}_{-0.07}$&$1.33^{+0.07}_{-0.07}$          &                                         \\
      && (14) &$N_{\rm{Swift/BAT}}$               &$2.24^{+0.78}_{-0.75}$&$1.87^{+0.46}_{-0.43}$          &                    \\
      &{- - -}& (15) &$N^{\rm{LOS}}_{\rm{H}}$               &$1.00\ (\rm{fixed})$&$<6.54$          &$10^{19}$ cm$^{-2}$  \\
      && (16) &$L_{2-10, \rm{E}}$               &$1.08$&$1.00$         &$10^{43}\ \rm{erg}\ \rm{s}^{-1}$   
       \\
\hline
{Mrk~739W} &{\tt zcutoffpl}& (17) &$\Gamma_{\rm{AGN}}$ &$1.55^{+0.54}_{-0.05(*)}$&$1.50^{+0.40}_{-0.00(*)}$    &        \\
       && (18) &$K_{\rm{AGN}}$        &$6.94^{+6.41}_{-1.31}$&$6.57^{+4.44}_{-0.95}$ 
      &$10^{-5}$ keV$^{-1}$ cm$^{-2}$ s$^{-1}$                 \\
      &{\tt pexmon}& (19)  &$R$            &$-0.8\ (\rm{fixed})$&\nodata    & \\
      &{\tt atable\{xclumpy\_v01\_RC(L).fits\}}& (20) &$N^{\rm{Equ}}_{\rm{H}}$          &\nodata&$3.76^{+1.74}_{-0.95}$ &$10^{23}$ cm$^{-2}$ \\
      &{- - -}& (21)  &$N^{\rm{LOS}}_{\rm{H}}$               &$7.26^{+4.48}_{-2.07}$&$6.89^{+3.18}_{-1.74}$          &$10^{21}$ cm$^{-2}$       \\
      && (22) &$L_{2-10, \rm{W}}$               &$7.32$&$7.46$         &$10^{41}\ \rm{erg}\ \rm{s}^{-1}$  
      \\
\hline
{Extended} &{\tt zcutoffpl}& (23) &$\Gamma_{\rm{Ext}}$        &$1.53^{+0.33}_{-0.03(*)}$&$1.50^{+0.34}_{-0.00(*)}$ &                 \\
      && (24) &$K_{\rm{Ext}}$                   &$2.31^{+0.38}_{-0.34}$&$2.30^{+0.38}_{-0.32}$           &$10^{-5}$ keV$^{-1}$ cm$^{-2}$ s$^{-1}$  \\
\hline
      &{\tt const0}& (25) &$C_{\rm{Chandra}}$         &$1.09\ (\rm{fixed})$&$1.09\ (\rm{fixed})$ &                                 \\
      && (26) &$C_{\rm{XMM}}$               &$0.89\ (\rm{fixed})$&$0.89\ (\rm{fixed})$         &                                         \\
      && (27) &$C_{\rm{Swift/BAT}}$               &$1.00\ (\rm{fixed})$&$1.00\ (\rm{fixed})$          &                                         \\
      &&      &$\chi^2/\rm{dof}$ $(\chi^2_r)$   &$649.0/618$ ($1.05$)&$644.7/618$ ($1.04$)&
\enddata
\tablecomments
{
(1) Hydrogen column density of warm absorbers at the epoch of the Chandra and XMM-Newton observation, respectively.
(2) Logarithmic ionization parameter, $\xi$ ($\rm erg~cm~s^{-1}$) of warm absorbers at the epoch of the Chandra and XMM-Newton observation, respectively. 
(3) Power-law photon index of the direct component.
(4) Power-law normalization at 1 keV.
(5) Normalization of the RELXILL model.
(6) The reflection strength of the pexrav model (the solid angle of the reflector normalized by $2\pi$).
(7) Normalization of the \textsf{zgauss} model.
(8) Hydrogen column density along the equatorial plane of the XCLUMPY torus model.
(9) Torus angular width.
(10) Normalization of the \textsf{apec} model.
(11) Temperature of the \textsf{apec} model.
(12)--(14) Time variability constant of Chandra, XMM-Newton, and Swift/BAT relative to NuSTAR, respectively.
(15) Hydrogen column density along the line of sight determined by the torus parameters through Equation~(\ref{N_H}). 
(16) Intrinsic 2--10 keV luminosity of Mrk~739E at the epoch of the Chandra observation.
(17)--(22) Parameters for Mrk~739W corresponding to (3), (4), (6), (8), (15) and (16), respectively. 
(23) Power-law photon index of the extended emission.
(24) Power-law normalization at 1 keV.
(25)--(27) Cross-calibration constant of Chandra, XMM-Newton, and Swift/BAT relative to NuSTAR, respectively.
}
\tablenotetext{*}{The parameter reaches a limit of its allowed range.}
\end{deluxetable*}

\begin{deluxetable}{ccc}
\tablewidth{\textwidth}

\tablecaption{Best-fit torus parameters of Mrk~739E \label{sigma-NH-chi}}
\tablehead{
\colhead{$\sigma$(degree)}      &
\colhead{$N^{\rm Equ}_{\rm H}$($10^{23}$ cm$^{-2}$)}      &
\colhead{$\chi^2/\rm{dof}$ $(\chi^2_r)$} 
}
\startdata
10 & $100^{+0.00(*)}_{-99.0(*)}$ & 644.0/618(1.04) \\
15 & $1.00^{+4.30}_{-0.00(*)}$ & 644.7/618(1.04)  \\
20 & $1.00^{+0.21}_{-0.00(*)}$ & 654.6/618(1.06)
\enddata

\tablecomments
{
$\sigma$ is a fixed value, while $N^{\rm Equ}_{\rm H}$ is a free parameter.
}
\tablenotetext{*}{The parameter reaches a limit of its allowed range.}

\end{deluxetable}

\section{DISCUSSION} \label{sec:discussion}

\subsection{Result Summary}

We have analyzed the highest-quality available broadband X-ray spectra
of the Mrk~739 system obtained with NuSTAR, Chandra, XMM-Newton, and
Swift/BAT, covering the 0.5--70 keV band.
Thanks to the superior angular
resolution of Chandra, it is possible to separately analyze the
spectra of three different regions, Mrk~739E, 
Mrk~739W,
and the extended emission (Figure~\ref{fig-ds9}).
We verify that
the spectra of the other instruments can be reproduced with the sum of
these three spectra by taking into account time variability in the flux of
the direct components and warm-absorber parameters (for Mrk~739E).
In particular, the NuSTAR data constrain well the broadband spectrum
of the brighter AGN, Mrk~739E. 
To model the reflection components from the tori, we
employ two models; \textsf{pexrav} (Model I, \citealt{Magdziarz1995}) and XCLUMPY (Model II, \citealt{Tanimoto_2019}).
Here, we mainly discuss the results obtained by Model II,
where more realistic torus geometry is assumed. 

Through this analysis, we are able to best constrain the X-ray
properties of the two AGNs. We obtain the intrinsic 2--10 keV
luminosities of $1.0\times 10^{43}$ and $7.5\times 10^{41}\ \rm{erg}\ \rm{s}^{-1}$ and absorption column densities by neutral material
of $N_{\rm{H}} < 6.5 \times 10^{19}\ \rm{cm}^{-2}$ and 
$N_{\rm{H}} =6.9^{+3.2}_{-1.7} \times 10^{21}\ \rm{cm}^{-2}$ for Mrk~739E and Mrk~739W, respectively, at the epoch of
the Chandra observation in April 2011. These results are essentially
consistent with the previous study using Chandra and Swift/BAT \citep{Koss_2011a}, except that
the column density of Mrk~739E is smaller than their estimate because of
the inclusion of the warm absorbers in our analysis. We confirm
that Mrk~739E is an unabsorbed moderately-luminous AGN, whereas Mrk
739W is a weakly absorbed, low-luminosity AGN. In the following, we
focus on the AGN properties of Mrk~739E, which is $\sim$15 times brighter than Mrk~739W in the 2--10 keV band.

\subsection{AGN Properties of Mrk~739E}

We examine whether the AGN torus in Mrk~739E has a larger covering
fraction 
than those of ``normal AGNs'' (i.e., those in nonmergers or early-stage
mergers) based on the results obtained by Model II.
The torus covering fraction, $C_{\rm{T}}^{(22)}$ is defined as the solid angle of torus matter with
column densities of $\log N_{\rm H}/{\rm{cm}^{-2}} >22$ normalized by
$4\pi$.
Using a Swift/BAT selected AGN sample, 
\citet{Ricci_2017b} find that the Eddington ratio ($\lambda_{\rm{Edd}}$)
is the key parameter that determines the torus structure; 
$C_{\rm{T}}^{(22)}$ rapidly decreases from $\sim$0.8 to $\sim$0.3
at $\log \lambda_{\rm{Edd}} > -1.5$. This suggests that the AGN tori
are regulated by radiation pressure on dusty gas within 
sphere of influence
of the SMBH (see also \citealt{Ricci2022}). 
By contrast, as we mention in Section~\ref{sec:intro},
\citet{Yamada_2021} find that AGNs in late-merger U/LIRGs show
$C_{\rm{T}}^{(22)} = 0.71\pm0.16$ (mean and standard deviation)
at $\log \lambda_{\rm{Edd}} > -1.5$, which is significantly larger than those of normal AGNs at
the same Eddington ratios (see also \citealt{Yamada_2020}). This implies that the AGNs in late mergers
have distinct structures from normal AGNs, possibly affected by the
merger-driven quasi-spherical mass outflow and resultant enhanced
outflow (see Figure~13 of \citealt{Yamada_2021}). 
Other recent studies also point towards a larger obscuration 
of AGNs in dual galaxy systems. \citet{DeRosa_2018} study X-ray
absorption properties of four dual systems with projected separations
of $30-60$ kpc, and find that the fraction of obscured ($\log N_{\rm
H}/{\rm{cm}^{-2}} >22$) AGNs, $84\pm 4\ \%$, is higher than that of
isolated AGNs detected with Swift/BAT ($46\pm 3\ \%$;
\citealt{Ricci2015}). \citet{Guainazzi_2021} investigate the X-ray
properties of 32 galaxy pairs with projected separations of $\leq$150
kpc and find that AGNs in galaxy pairs tend to be obscured (a fraction
of $>70\%$).

Using the torus parameters ($\sigma$ and $N^{\rm{Equ}}_{\rm{H}}$) of
XCLUMPY, we calculate $C_{\rm{T}}^{(22)}$ as follows (see also Equation (6) in \citealt{Ogawa_2021}). 
Here, we define the critical elevation angle $\theta_{\rm{c}}$, which satisfies $N_{\rm{H}}^{\rm{LOS}}(\theta_{\rm{c}})=10^{22}$ in the  Equation~(\ref{N_H}).
Then, $C_{\rm{T}}^{(22)}$ can be calculated as follows: 
\begin{eqnarray} \label{C_T}
C_{\rm T}^{(22)} 
&=& \frac{1}{4\pi}\int_{0}^{2\pi}
\int_{\frac{\pi}{2}-\theta_{\rm{c}}}^{\frac{\pi}{2}+\theta_{\rm{c}}} \sin \theta d\theta d\phi
\nonumber\\
&=& \rm{sin}
\Biggl(
\sigma
\sqrt{
\rm{ln}\biggl(
\frac{\textit{N}^{\, \rm{Equ}}_{\, \rm{H}}}{10^{22}\ \rm{cm}^{-2}}
\biggr)
}
\Biggr)
\end{eqnarray}
Since we are not able to determine $N^{\rm Equ}_{\rm H}$ and $\sigma$
simultaneously due to their parameter coupling, we estimate $N^{\rm
Equ}_{\rm H}$ for the three cases, $\sigma = 10^{\circ}$, $15^{\circ}$, or $20^{\circ}$ (see Section~\ref{subsec:model}).
We find that $C_{\rm{T}}^{(22)}$ of Mrk~739E is smaller than 0.50 at a 90\% confidence limit for any values of $\sigma$ we adopted.

\begin{figure}
\centering
\includegraphics[keepaspectratio,scale=0.55]
{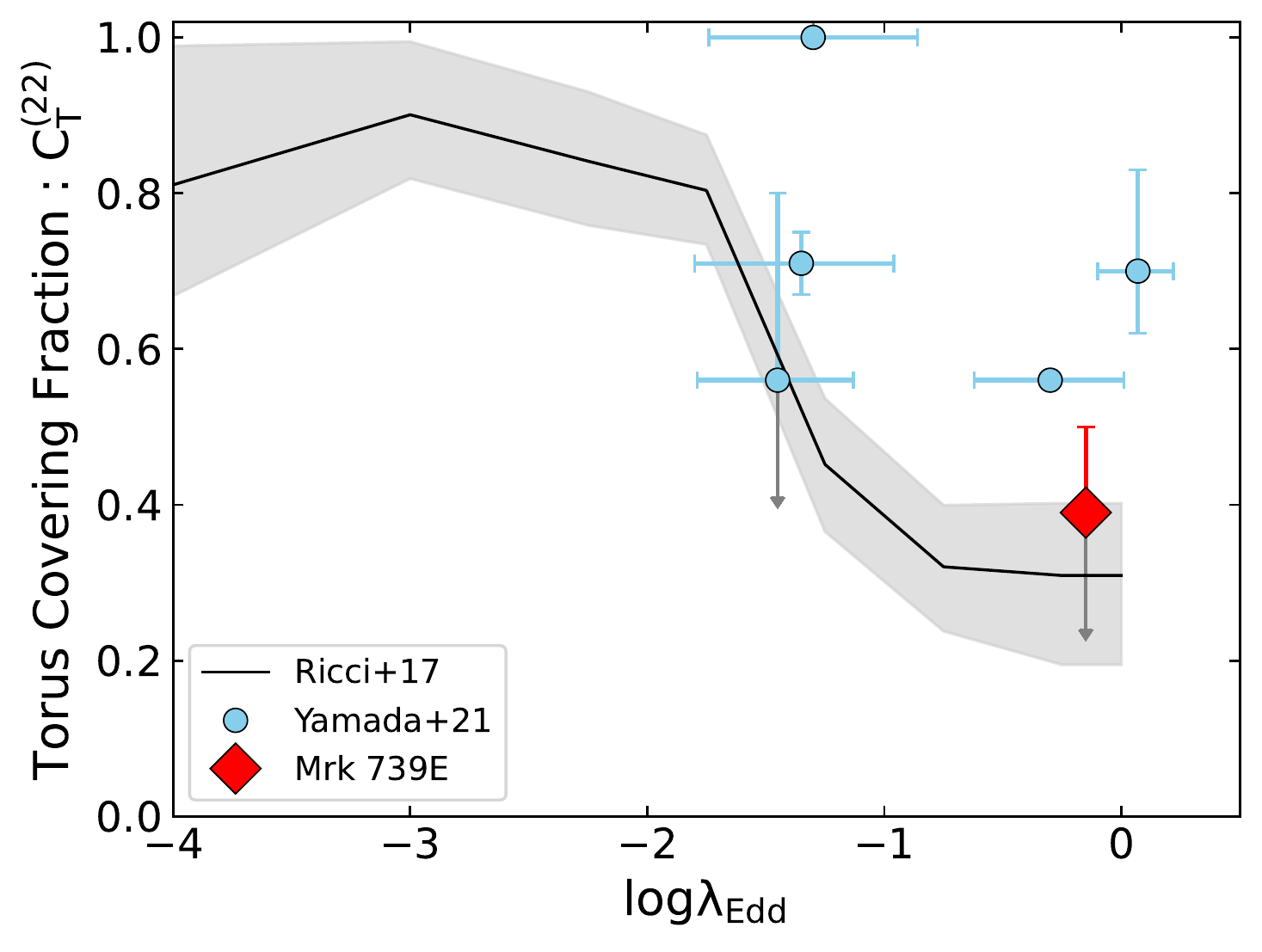}
\caption{
  Relationship between $\lambda_{\rm{Edd}}$ and $C_{\rm{T}}^{(22)}$ for Swift/BAT selected AGNs (black line) and 1$\sigma$ uncertainties (the gray shaded areas) obtained by \citet{Ricci_2017b}. Our result for Mrk~739E (red diamond; the best-fit value obtained for $\sigma=15^\circ$ is adopted) and those for the late-merger U/LIRGs 
  (cyan circles; \citealt{Yamada_2021}) are overplotted.
  The arrows represent the results reaching the lower boundary.
}
\label{Ricci_diagram}
\end{figure}

In the Figure~\ref{Ricci_diagram}, the relation between
$C_{\rm{T}}^{(22)}$ and $\lambda_{\rm{Edd}}$ obtained by
\citet{Ricci_2017b} for local hard X-ray selected AGNs, which are typically found in nonmerging galaxies, is shown. 
The points correspond to
Mrk~739E (red) and the late-merger U/LIRGs (cyan; \citealt{Yamada_2021}).
 The $C_{\rm{T}}^{(22)}$ values of the late-merger U/LIRGs are
 calculated from the XCLUMPY parameters in the same manner as in this paper.
Here, the Eddington ratio of Mrk~739E ($\log{\lambda_{\rm Edd}} =
-0.15$) 
is calculated from the bolometric luminosity based on the UV data
\citep{Koss_2011a}, $L_{\rm{bol}}=1.0 \times 10^{45}\ \rm{erg}\ \rm{s}^{-1}$, and the black hole mass estimated by \citet{Koss_2011a}, $\log M_{\rm{BH}}/M_{\odot} \sim 7.04$.

Figure~\ref{Ricci_diagram} shows that the AGN torus of Mrk~739E
has a smaller covering fraction
than those of the late-merger U/LIRGs. Interestingly, the covering fraction is
consistent with those of nonmergers at the same Eddington ratios.
To verify the statistical significance of this result, we conduct Monte Carlo simulations with a sample size of $10^6$ by assuming that the intrinsic values of $C_{\rm{T}}^{(22)}$ follow a gaussian distribution with a mean of $0.71$ and a standard deviation of $0.16$, the distribution of the late-merger U/LIRGs.
Additionally, we assume that $C_{\rm{T}}^{(22)}$ varies from its intrinsic value with a standard deviation of $0.07$,
the 1-$\sigma$ error obtained by our analysis.
We find that the probability of having $C_{\rm{T}}^{(22)}$ smaller than $0.39$, the best fit value of our analysis,
is $3.5\%$. We thus suggest that 
the AGN torus of Mrk~739E has a smaller covering fraction than those of the late-merger U/LIRGs at a $>$95\% confidence level.
Accordingly, we suggest that Mrk~739E is an exception to the idea that dual systems generally show large obscurations (e.g., \citealt{DeRosa_2018}, \citealt{Guainazzi_2021}).
The bolometric to 2--10 keV luminosity correction factor of Mrk~739E
is calculated to be $\kappa_{2-10} \sim 90$ (based on the Chandra
spectrum).
Considering the fact that Mrk~739E has a high Eddington ratio
($\log{\lambda_{\rm Edd}} = -0.15$),
this $\kappa_{2-10}$ value is consistent with typical values of
local hard X-ray selected AGNs at similar Eddington ratios, which
 show higher bolometric correction factors compared with those of
 AGNs at lower Eddington ratios ($\kappa_{2-10} \sim 20$ at $\log
 \lambda_{\rm Edd} < -1.0$, see e.g., \citealt{Vasudevan_Fabian_2007};
 \citealt{Lusso2012}).  However, it is much smaller than those found
for the late-merger U/LIRGs ($\kappa_{2-10} \sim 1000$;
\citealt{Yamada_2021}). This also supports the idea that the AGN in
Mrk~739E is just neither deeply buried nor X-ray weak, even though it
is in a late merger.

The difference in the AGN structure between Mrk~739E and late-merger
U/LIRGs may be explained by considering the properties of its host galaxy. The stellar
mass of Mrk~739E is estimated to be $\log{M_{\rm{stellar}}/M_{\odot}} \sim 10.86$ based on the
K-band luminosity excluding the AGN region \citep{Tubin_2021}. Using the far-infrared
luminosity of the total Mrk~739 system, the SFR of Mrk~739E is
constrained to be lower than $6.9 M_{\odot}\ \rm{yr}^{-1}$,
which includes the contribution from
Mrk~739W. These $M_{\rm{stellar}}$ and SFR values locates Mrk~739E as a main
sequence star-forming galaxy or a star-formation quenching galaxy
\citep{Tubin_2021}, in contrast to the late-merger U/LIRGs, which are mostly
star burst galaxies (e.g., \citealt{Shangguan_2019}; \citealt{Yamada_2021}). Assuming that the SFR is
proportional to the 1.4th power of gas mass \citep{Schmidt_1959},
$\sim$10
times smaller SFR-$M_{\rm{stellar}}$ ratio of Mrk~739E means that it has a
$\sim$0.2
times smaller gas-to-stellar mass ratio compared with the
late-merger U/LIRGs. Thus, our results suggest that the gas-to-mass
ratio of the host galaxy is an important parameter to determine the
circumnuclear environment of an AGN in a late merger. In fact, recent
numerical simulations of galaxy mergers show that both SFR and nuclear
obscuration (in terms of column density) become weaker in gas-poor
mergers than in gas-rich ones \citep{Blecha_2018}. Our results are consistent with
this picture.

\section{CONCLUSION} \label{sec:conclusion}

In this paper, we have reported the results of broadband X-ray
spectral (0.5--70 keV) analysis of the dual-AGN system Mrk~739,
utilizing currently the highest-quality data observed with NuSTAR,
Chandra, XMM-Newton and Swift/BAT. The latest X-ray clumpy torus model,
XCLUMPY \citep{Tanimoto_2019}, has been applied. Our main conclusions are summarized
below.

\begin{enumerate}

\item The X-ray
spectrum of Mrk~739E is well reproduced by a direct
power-law component and its reflection components from the accretion
disk (RELXILL) and the torus (XCLUMPY). Time-variable warm absorbers
are detected. The X-ray spectrum of Mrk~739W can be modelled by an absorbed power-law
and its reflection from the torus. 

\item We have estimated the line of sight column densities by cold
matter to be $N_{\rm{H}} < 6.5 \times 10^{19}\ \rm{cm}^{-2}$ and 
$N_{\rm{H}} =6.9^{+3.2}_{-1.7} \times 10^{21}\ \rm{cm}^{-2}$
and the intrinsic X-ray luminosities to be 
$1.0\times 10^{43}$ and
$7.5\times 10^{41}\ \rm{erg}\ \rm{s}^{-1}$ for
Mrk~739E and Mrk~739W, respectively. This confirms that Mrk
739E is an unabsorbed moderately-luminous AGN, whereas Mrk~739W is a
weakly absorbed, low-luminosity AGN.

\item On the basis of the parameters of XCLUMPY, we have determined
the torus covering fraction of Mrk~739E to be $C_{\rm{T}}^{(22)}
< 0.50$. 
This is smaller than those of local late-merger
U/LIRGs at similar Eddington rates, $C_{\rm{T}}^{(22)} = 0.71\pm0.16$
(mean and standard deviation; \citealt{Yamada_2021}).

\item Our result is a counterexample against the idea that AGNs are
always deeply buried by gas and dust (i.e., the AGN tori have large
covering fractions) in the late-stage mergers. We suggest that the
gas-to-mass ratio of the host galaxy is an important parameter to
determine the circumnuclear environment, which is consistent with
recent numerical simulations of galaxy mergers.
Further studies based on a larger sample of this type of AGNs are required to confirm this conclusion.

\end{enumerate}

\begin{acknowledgements}

This work has been financially supported by the Grant-in-Aid for JSPS
Research Fellowships 19J22216, 22K20391 (S.Y.), 20J00119 (A.T.), 21J13894 (S.O.), and for Scientific Research 20H01946 (Y.U.).
S.Y. is also grateful for support from RIKEN Special Postdoctoral Researcher Program.
C.R. acknowledges support from the Fondecyt Iniciacion grant 11190831 and ANID BASAL project FB210003. 

We have made use of NuSTAR Data Analysis Software (NuSTARDAS) jointly developed by the Space Science Data Center (SSDC; ASI, Italy) and the California Institute of Technology (USA).
We have utilized data from Chandra, supported by the Chandra X-ray Observatory Center operated by the Smithsonian Astrophysical Observatory
and data from XMM-Newton, an ESA science mission with instruments and contributions directly funded by ESA Member States and NASA. We also acknowledge the use of public data from the Swift data archive.
\end{acknowledgements}

\facilities{NuSTAR, Chandra, XMM-Newton, Swift.} 

\software{HEAsoft (v6.28; \citealt{heasoft}), NuSTARDAS (v2.0.0), SAS (v18.0.0; \citealt{Gabriel_2004}), CIAO (v4.12; \citealt{Fruscione_2006}), XSPEC (v12.11.1; \citealt{Arnaud_1996}), XCLUMPY \citep{Tanimoto_2019}, RELXILL (v1.4.0; \citealt{Dauser_2013}; \citealt{Garc_a_2014}). }

\bibliography{sample631}{}

\begin{thebibliography}{}
\expandafter\ifx\csname natexlab\endcsname\relax\def\natexlab#1{#1}\fi
\providecommand{\url}[1]{\href{#1}{#1}}
\providecommand{\dodoi}[1]{doi:~\href{http://doi.org/#1}{\nolinkurl{#1}}}
\providecommand{\doeprint}[1]{\href{http://ascl.net/#1}{\nolinkurl{http://ascl.net/#1}}}
\providecommand{\doarXiv}[1]{\href{https://arxiv.org/abs/#1}{\nolinkurl{https://arxiv.org/abs/#1}}}

\bibitem[{{Anders} \& {Grevesse}(1989)}]{Anders_1989}
{Anders}, E., \& {Grevesse}, N. 1989, \gca, 53, 197,
  \dodoi{10.1016/0016-7037(89)90286-X}

\bibitem[{{Arnaud}(1996)}]{Arnaud_1996}
{Arnaud}, K.~A. 1996, in Astronomical Society of the Pacific Conference Series,
  Vol. 101, Astronomical Data Analysis Software and Systems V, ed. G.~H.
  {Jacoby} \& J.~{Barnes}, 17

\bibitem[{{Bautista} \& {Kallman}(2001)}]{Bautista_Kallman_2001}
{Bautista}, M.~A., \& {Kallman}, T.~R. 2001, \apjs, 134, 139,
  \dodoi{10.1086/320363}

\bibitem[{{Blecha} {et~al.}(2018){Blecha}, {Snyder}, {Satyapal}, \&
  {Ellison}}]{Blecha_2018}
{Blecha}, L., {Snyder}, G.~F., {Satyapal}, S., \& {Ellison}, S.~L. 2018,
  \mnras, 478, 3056, \dodoi{10.1093/mnras/sty1274}

\bibitem[{Dauser {et~al.}(2013)Dauser, Garcia, Wilms, Böck, Brenneman,
  Falanga, Fukumura, \& Reynolds}]{Dauser_2013}
Dauser, T., Garcia, J., Wilms, J., {et~al.} 2013, Monthly Notices of the Royal
  Astronomical Society, 430, 1694–1708, \dodoi{10.1093/mnras/sts710}

\bibitem[{Dauser {et~al.}(2010)Dauser, Wilms, Reynolds, \&
  Brenneman}]{Dauser_2010}
Dauser, T., Wilms, J., Reynolds, C.~S., \& Brenneman, L.~W. 2010, Monthly
  Notices of the Royal Astronomical Society, 409, 1534–1540,
  \dodoi{10.1111/j.1365-2966.2010.17393.x}

\bibitem[{{De Rosa} {et~al.}(2018){De Rosa}, {Vignali}, {Husemann}, {Bianchi},
  {Bogdanovi{\'c}}, {Guainazzi}, {Herrero-Illana}, {Komossa}, {Kun}, {Loiseau},
  {Paragi}, {Perez-Torres}, \& {Piconcelli}}]{DeRosa_2018}
{De Rosa}, A., {Vignali}, C., {Husemann}, B., {et~al.} 2018, \mnras, 480, 1639,
  \dodoi{10.1093/mnras/sty1867}

\bibitem[{{Ferrarese} \& {Merritt}(2000)}]{Ferrarese_2000}
{Ferrarese}, L., \& {Merritt}, D. 2000, \apjl, 539, L9, \dodoi{10.1086/312838}

\bibitem[{{Fruscione} {et~al.}(2006){Fruscione}, {McDowell}, {Allen},
  {Brickhouse}, {Burke}, {Davis}, {Durham}, {Elvis}, {Galle}, {Harris},
  {Huenemoerder}, {Houck}, {Ishibashi}, {Karovska}, {Nicastro}, {Noble},
  {Nowak}, {Primini}, {Siemiginowska}, {Smith}, \& {Wise}}]{Fruscione_2006}
{Fruscione}, A., {McDowell}, J.~C., {Allen}, G.~E., {et~al.} 2006, in Society
  of Photo-Optical Instrumentation Engineers (SPIE) Conference Series, Vol.
  6270, Society of Photo-Optical Instrumentation Engineers (SPIE) Conference
  Series, ed. D.~R. {Silva} \& R.~E. {Doxsey}, 62701V,
  \dodoi{10.1117/12.671760}

\bibitem[{{Gabriel} {et~al.}(2004){Gabriel}, {Denby}, {Fyfe}, {Hoar}, {Ibarra},
  {Ojero}, {Osborne}, {Saxton}, {Lammers}, \& {Vacanti}}]{Gabriel_2004}
{Gabriel}, C., {Denby}, M., {Fyfe}, D.~J., {et~al.} 2004, in Astronomical
  Society of the Pacific Conference Series, Vol. 314, Astronomical Data
  Analysis Software and Systems (ADASS) XIII, ed. F.~{Ochsenbein}, M.~G.
  {Allen}, \& D.~{Egret}, 759

\bibitem[{García {et~al.}(2014)García, Dauser, Lohfink, Kallman, Steiner,
  McClintock, Brenneman, Wilms, Eikmann, Reynolds, \& et~al.}]{Garc_a_2014}
García, J., Dauser, T., Lohfink, A., {et~al.} 2014, The Astrophysical Journal,
  782, 76, \dodoi{10.1088/0004-637x/782/2/76}

\bibitem[{{Garmire} {et~al.}(2003){Garmire}, {Bautz}, {Ford}, {Nousek}, \&
  {Ricker}}]{Garmire_2003}
{Garmire}, G.~P., {Bautz}, M.~W., {Ford}, P.~G., {Nousek}, J.~A., \& {Ricker},
  George~R., J. 2003, in Society of Photo-Optical Instrumentation Engineers
  (SPIE) Conference Series, Vol. 4851, X-Ray and Gamma-Ray Telescopes and
  Instruments for Astronomy., ed. J.~E. {Truemper} \& H.~D. {Tananbaum},
  28--44, \dodoi{10.1117/12.461599}

\bibitem[{{Gehrels} {et~al.}(2004){Gehrels}, {Chincarini}, {Giommi}, {Mason},
  {Nousek}, {Wells}, {White}, {Barthelmy}, {Burrows}, {Cominsky}, {Hurley},
  {Marshall}, {M{\'e}sz{\'a}ros}, {Roming}, {Angelini}, {Barbier}, {Belloni},
  {Campana}, {Caraveo}, {Chester}, {Citterio}, {Cline}, {Cropper}, {Cummings},
  {Dean}, {Feigelson}, {Fenimore}, {Frail}, {Fruchter}, {Garmire}, {Gendreau},
  {Ghisellini}, {Greiner}, {Hill}, {Hunsberger}, {Krimm}, {Kulkarni}, {Kumar},
  {Lebrun}, {Lloyd-Ronning}, {Markwardt}, {Mattson}, {Mushotzky}, {Norris},
  {Osborne}, {Paczynski}, {Palmer}, {Park}, {Parsons}, {Paul}, {Rees},
  {Reynolds}, {Rhoads}, {Sasseen}, {Schaefer}, {Short}, {Smale}, {Smith},
  {Stella}, {Tagliaferri}, {Takahashi}, {Tashiro}, {Townsley}, {Tueller},
  {Turner}, {Vietri}, {Voges}, {Ward}, {Willingale}, {Zerbi}, \&
  {Zhang}}]{Gehrels_2004}
{Gehrels}, N., {Chincarini}, G., {Giommi}, P., {et~al.} 2004, \apj, 611, 1005,
  \dodoi{10.1086/422091}

\bibitem[{{Guainazzi} {et~al.}(2021){Guainazzi}, {De Rosa}, {Bianchi},
  {Husemann}, {Bogdanovic}, {Komossa}, {Loiseau}, {Paragi}, {P{\'e}rez-Torres},
  {Piconcelli}, \& {Vignali}}]{Guainazzi_2021}
{Guainazzi}, M., {De Rosa}, A., {Bianchi}, S., {et~al.} 2021, \mnras, 504, 393,
  \dodoi{10.1093/mnras/stab808}

\bibitem[{Harrison {et~al.}(2013)Harrison, Craig, Christensen, Hailey, Zhang,
  Boggs, Stern, Cook, Forster, Giommi, \& et~al.}]{Harrison_2013}
Harrison, F.~A., Craig, W.~W., Christensen, F.~E., {et~al.} 2013, The
  Astrophysical Journal, 770, 103, \dodoi{10.1088/0004-637x/770/2/103}

\bibitem[{Hopkins {et~al.}(2008)Hopkins, Hernquist, Cox, \&
  Kereš}]{Hopkins_2008}
Hopkins, P.~F., Hernquist, L., Cox, T.~J., \& Kereš, D. 2008, The
  Astrophysical Journal Supplement Series, 175, 356–389,
  \dodoi{10.1086/524362}

\bibitem[{{Hopkins} {et~al.}(2006){Hopkins}, {Somerville}, {Hernquist}, {Cox},
  {Robertson}, \& {Li}}]{Hopkins_2006}
{Hopkins}, P.~F., {Somerville}, R.~S., {Hernquist}, L., {et~al.} 2006, \apj,
  652, 864, \dodoi{10.1086/508503}

\bibitem[{{Imanishi} {et~al.}(2006){Imanishi}, {Dudley}, \&
  {Maloney}}]{Imanishi_2006}
{Imanishi}, M., {Dudley}, C.~C., \& {Maloney}, P.~R. 2006, \apj, 637, 114,
  \dodoi{10.1086/498391}

\bibitem[{{Imanishi} {et~al.}(2008){Imanishi}, {Nakagawa}, {Ohyama},
  {Shirahata}, {Wada}, {Onaka}, \& {Oi}}]{Imanishi_2008}
{Imanishi}, M., {Nakagawa}, T., {Ohyama}, Y., {et~al.} 2008, \pasj, 60, S489,
  \dodoi{10.1093/pasj/60.sp2.S489}

\bibitem[{{Imanishi} \& {Saito}(2014)}]{Imanishi_2014}
{Imanishi}, M., \& {Saito}, Y. 2014, \apj, 780, 106,
  \dodoi{10.1088/0004-637X/780/1/106}

\bibitem[{{Jansen} {et~al.}(2001){Jansen}, {Lumb}, {Altieri}, {Clavel}, {Ehle},
  {Erd}, {Gabriel}, {Guainazzi}, {Gondoin}, {Much}, {Munoz}, {Santos},
  {Schartel}, {Texier}, \& {Vacanti}}]{Jansen_2001}
{Jansen}, F., {Lumb}, D., {Altieri}, B., {et~al.} 2001, \aap, 365, L1,
  \dodoi{10.1051/0004-6361:20000036}

\bibitem[{{Kallman} \& {Bautista}(2001)}]{Kallman_Bautista_2001}
{Kallman}, T., \& {Bautista}, M. 2001, \apjs, 133, 221, \dodoi{10.1086/319184}

\bibitem[{{Kawaguchi} {et~al.}(2020){Kawaguchi}, {Yutani}, \&
  {Wada}}]{Kawaguchi_2020}
{Kawaguchi}, T., {Yutani}, N., \& {Wada}, K. 2020, \apj, 890, 125,
  \dodoi{10.3847/1538-4357/ab655a}

\bibitem[{{Kawamuro} {et~al.}(2016){Kawamuro}, {Ueda}, {Tazaki}, {Ricci}, \&
  {Terashima}}]{Kawamuro_2016}
{Kawamuro}, T., {Ueda}, Y., {Tazaki}, F., {Ricci}, C., \& {Terashima}, Y. 2016,
  \apjs, 225, 14, \dodoi{10.3847/0067-0049/225/1/14}

\bibitem[{Kormendy \& Ho(2013)}]{Kormendy_2013}
Kormendy, J., \& Ho, L.~C. 2013, Annual Review of Astronomy and Astrophysics,
  51, 511–653, \dodoi{10.1146/annurev-astro-082708-101811}

\bibitem[{Koss {et~al.}(2012)Koss, Mushotzky, Treister, Veilleux, Vasudevan, \&
  Trippe}]{Koss_2012}
Koss, M., Mushotzky, R., Treister, E., {et~al.} 2012, The Astrophysical
  Journal, 746, L22, \dodoi{10.1088/2041-8205/746/2/l22}

\bibitem[{Koss {et~al.}(2011)Koss, Mushotzky, Veilleux, Vasudevan, Miller,
  Sanders, Schawinski, \& Trippe}]{Koss_2011a}
Koss, M., Mushotzky, R., Veilleux, S., {et~al.} 2011, The Astrophysical
  Journal, 735, L42, \dodoi{10.1088/2041-8205/735/2/l42}

\bibitem[{{Lusso} {et~al.}(2012){Lusso}, {Comastri}, {Simmons}, {Mignoli},
  {Zamorani}, {Vignali}, {Brusa}, {Shankar}, {Lutz}, {Trump}, {Maiolino},
  {Gilli}, {Bolzonella}, {Puccetti}, {Salvato}, {Impey}, {Civano}, {Elvis},
  {Mainieri}, {Silverman}, {Koekemoer}, {Bongiorno}, {Merloni}, {Berta}, {Le
  Floc'h}, {Magnelli}, {Pozzi}, \& {Riguccini}}]{Lusso2012}
{Lusso}, E., {Comastri}, A., {Simmons}, B.~D., {et~al.} 2012, \mnras, 425, 623,
  \dodoi{10.1111/j.1365-2966.2012.21513.x}

\bibitem[{{Madsen} {et~al.}(2017){Madsen}, {Beardmore}, {Forster}, {Guainazzi},
  {Marshall}, {Miller}, {Page}, \& {Stuhlinger}}]{Madsen_2017}
{Madsen}, K.~K., {Beardmore}, A.~P., {Forster}, K., {et~al.} 2017, \aj, 153, 2,
  \dodoi{10.3847/1538-3881/153/1/2}

\bibitem[{{Magdziarz} \& {Zdziarski}(1995)}]{Magdziarz1995}
{Magdziarz}, P., \& {Zdziarski}, A.~A. 1995, \mnras, 273, 837,
  \dodoi{10.1093/mnras/273.3.837}

\bibitem[{{Magorrian} {et~al.}(1998){Magorrian}, {Tremaine}, {Richstone},
  {Bender}, {Bower}, {Dressler}, {Faber}, {Gebhardt}, {Green}, {Grillmair},
  {Kormendy}, \& {Lauer}}]{Magorrian_1998}
{Magorrian}, J., {Tremaine}, S., {Richstone}, D., {et~al.} 1998, \aj, 115,
  2285, \dodoi{10.1086/300353}

\bibitem[{{Nandra} {et~al.}(2007){Nandra}, {O'Neill}, {George}, \&
  {Reeves}}]{Nandra_2007}
{Nandra}, K., {O'Neill}, P.~M., {George}, I.~M., \& {Reeves}, J.~N. 2007,
  \mnras, 382, 194, \dodoi{10.1111/j.1365-2966.2007.12331.x}

\bibitem[{{Nasa High Energy Astrophysics Science Archive Research Center
  (Heasarc)}(2014)}]{heasoft}
{Nasa High Energy Astrophysics Science Archive Research Center (Heasarc)}.
  2014, {HEAsoft: Unified Release of FTOOLS and XANADU}, Astrophysics Source
  Code Library, record ascl:1408.004.
\newblock \doeprint{1408.004}

\bibitem[{{Netzer} {et~al.}(1987){Netzer}, {Kollatschny}, \&
  {Fricke}}]{Netzer_1987}
{Netzer}, H., {Kollatschny}, W., \& {Fricke}, K.~J. 1987, \aap, 171, 41

\bibitem[{{Ogawa} {et~al.}(2021){Ogawa}, {Ueda}, {Tanimoto}, \&
  {Yamada}}]{Ogawa_2021}
{Ogawa}, S., {Ueda}, Y., {Tanimoto}, A., \& {Yamada}, S. 2021, \apj, 906, 84,
  \dodoi{10.3847/1538-4357/abccce}

\bibitem[{{Ogawa} {et~al.}(2019){Ogawa}, {Ueda}, {Yamada}, {Tanimoto}, \&
  {Kawaguchi}}]{Ogawa_2019}
{Ogawa}, S., {Ueda}, Y., {Yamada}, S., {Tanimoto}, A., \& {Kawaguchi}, T. 2019,
  \apj, 875, 115, \dodoi{10.3847/1538-4357/ab0e08}

\bibitem[{{Oh} {et~al.}(2018){Oh}, {Koss}, {Markwardt}, {Schawinski},
  {Baumgartner}, {Barthelmy}, {Cenko}, {Gehrels}, {Mushotzky}, {Petulante},
  {Ricci}, {Lien}, \& {Trakhtenbrot}}]{Oh_2018}
{Oh}, K., {Koss}, M., {Markwardt}, C.~B., {et~al.} 2018, \apjs, 235, 4,
  \dodoi{10.3847/1538-4365/aaa7fd}

\bibitem[{{Patrick} {et~al.}(2012){Patrick}, {Reeves}, {Porquet}, {Markowitz},
  {Braito}, \& {Lobban}}]{Patrick_2012}
{Patrick}, A.~R., {Reeves}, J.~N., {Porquet}, D., {et~al.} 2012, \mnras, 426,
  2522, \dodoi{10.1111/j.1365-2966.2012.21868.x}

\bibitem[{{Ramos Almeida} \& {Ricci}(2017)}]{Ramos_2017}
{Ramos Almeida}, C., \& {Ricci}, C. 2017, Nature Astronomy, 1, 679,
  \dodoi{10.1038/s41550-017-0232-z}

\bibitem[{{Ricci} {et~al.}(2015){Ricci}, {Ueda}, {Koss}, {Trakhtenbrot},
  {Bauer}, \& {Gandhi}}]{Ricci2015}
{Ricci}, C., {Ueda}, Y., {Koss}, M.~J., {et~al.} 2015, \apjl, 815, L13,
  \dodoi{10.1088/2041-8205/815/1/L13}

\bibitem[{{Ricci} {et~al.}(2017{\natexlab{a}}){Ricci}, {Bauer}, {Treister},
  {Schawinski}, {Privon}, {Blecha}, {Arevalo}, {Armus}, {Harrison}, {Ho},
  {Iwasawa}, {Sanders}, \& {Stern}}]{Ricci_2017a}
{Ricci}, C., {Bauer}, F.~E., {Treister}, E., {et~al.} 2017{\natexlab{a}},
  \mnras, 468, 1273, \dodoi{10.1093/mnras/stx173}

\bibitem[{{Ricci} {et~al.}(2017{\natexlab{b}}){Ricci}, {Trakhtenbrot}, {Koss},
  {Ueda}, {Schawinski}, {Oh}, {Lamperti}, {Mushotzky}, {Treister}, {Ho},
  {Weigel}, {Bauer}, {Paltani}, {Fabian}, {Xie}, \& {Gehrels}}]{Ricci_2017b}
{Ricci}, C., {Trakhtenbrot}, B., {Koss}, M.~J., {et~al.} 2017{\natexlab{b}},
  \nat, 549, 488, \dodoi{10.1038/nature23906}

\bibitem[{{Ricci} {et~al.}(2018){Ricci}, {Ho}, {Fabian}, {Trakhtenbrot},
  {Koss}, {Ueda}, {Lohfink}, {Shimizu}, {Bauer}, {Mushotzky}, {Schawinski},
  {Paltani}, {Lamperti}, {Treister}, \& {Oh}}]{Ricci_2018}
{Ricci}, C., {Ho}, L.~C., {Fabian}, A.~C., {et~al.} 2018, \mnras, 480, 1819,
  \dodoi{10.1093/mnras/sty1879}

\bibitem[{{Ricci} {et~al.}(2021){Ricci}, {Privon}, {Pfeifle}, {Armus},
  {Iwasawa}, {Torres-Alb{\`a}}, {Satyapal}, {Bauer}, {Treister}, {Ho}, {Aalto},
  {Ar{\'e}valo}, {Barcos-Mu{\~n}oz}, {Charmandaris}, {Diaz-Santos}, {Evans},
  {Gao}, {Inami}, {Koss}, {Lansbury}, {Linden}, {Medling}, {Sanders}, {Song},
  {Stern}, {U}, {Ueda}, \& {Yamada}}]{Ricci_2021}
{Ricci}, C., {Privon}, G.~C., {Pfeifle}, R.~W., {et~al.} 2021, \mnras, 506,
  5935, \dodoi{10.1093/mnras/stab2052}

\bibitem[{{Ricci} {et~al.}(2022){Ricci}, {Ananna}, {Temple}, {Urry}, {Koss},
  {Trakhtenbrot}, {Ueda}, {Stern}, {Bauer}, {Treister}, {Privon}, {Oh},
  {Paltani}, {Stalevski}, {Ho}, {Fabian}, {Mushotzky}, {Chang}, {Ricci},
  {Kakkad}, {Sartori}, {Baer}, {Caglar}, {Powell}, \& {Harrison}}]{Ricci2022}
{Ricci}, C., {Ananna}, T.~T., {Temple}, M.~J., {et~al.} 2022, arXiv e-prints,
  arXiv:2209.00014.
\newblock \doarXiv{2209.00014}

\bibitem[{{Sanders} \& {Mirabel}(1996)}]{Sanders_1996}
{Sanders}, D.~B., \& {Mirabel}, I.~F. 1996, \araa, 34, 749,
  \dodoi{10.1146/annurev.astro.34.1.749}

\bibitem[{{Schmidt}(1959)}]{Schmidt_1959}
{Schmidt}, M. 1959, \apj, 129, 243, \dodoi{10.1086/146614}

\bibitem[{{Shangguan} {et~al.}(2019){Shangguan}, {Ho}, {Li}, {Zhuang}, {Xie},
  \& {Li}}]{Shangguan_2019}
{Shangguan}, J., {Ho}, L.~C., {Li}, R., {et~al.} 2019, \apj, 870, 104,
  \dodoi{10.3847/1538-4357/aaf21a}

\bibitem[{{Stierwalt} {et~al.}(2013){Stierwalt}, {Armus}, {Surace}, {Inami},
  {Petric}, {Diaz-Santos}, {Haan}, {Charmandaris}, {Howell}, {Kim}, {Marshall},
  {Mazzarella}, {Spoon}, {Veilleux}, {Evans}, {Sanders}, {Appleton}, {Bothun},
  {Bridge}, {Chan}, {Frayer}, {Iwasawa}, {Kewley}, {Lord}, {Madore},
  {Melbourne}, {Murphy}, {Rich}, {Schulz}, {Sturm}, {Vavilkin}, \&
  {Xu}}]{Stierwalt_2013}
{Stierwalt}, S., {Armus}, L., {Surace}, J.~A., {et~al.} 2013, \apjs, 206, 1,
  \dodoi{10.1088/0067-0049/206/1/1}

\bibitem[{{Str{\"u}der} {et~al.}(2001){Str{\"u}der}, {Briel}, {Dennerl},
  {Hartmann}, {Kendziorra}, {Meidinger}, {Pfeffermann}, {Reppin}, {Aschenbach},
  {Bornemann}, {Br{\"a}uninger}, {Burkert}, {Elender}, {Freyberg}, {Haberl},
  {Hartner}, {Heuschmann}, {Hippmann}, {Kastelic}, {Kemmer}, {Kettenring},
  {Kink}, {Krause}, {M{\"u}ller}, {Oppitz}, {Pietsch}, {Popp}, {Predehl},
  {Read}, {Stephan}, {St{\"o}tter}, {Tr{\"u}mper}, {Holl}, {Kemmer}, {Soltau},
  {St{\"o}tter}, {Weber}, {Weichert}, {von Zanthier}, {Carathanassis}, {Lutz},
  {Richter}, {Solc}, {B{\"o}ttcher}, {Kuster}, {Staubert}, {Abbey}, {Holland},
  {Turner}, {Balasini}, {Bignami}, {La Palombara}, {Villa}, {Buttler},
  {Gianini}, {Lain{\'e}}, {Lumb}, \& {Dhez}}]{Struder_2001}
{Str{\"u}der}, L., {Briel}, U., {Dennerl}, K., {et~al.} 2001, \aap, 365, L18,
  \dodoi{10.1051/0004-6361:20000066}

\bibitem[{Tanimoto {et~al.}(2019)Tanimoto, Ueda, Odaka, Kawaguchi, Fukazawa, \&
  Kawamuro}]{Tanimoto_2019}
Tanimoto, A., Ueda, Y., Odaka, H., {et~al.} 2019, The Astrophysical Journal,
  877, 95, \dodoi{10.3847/1538-4357/ab1b20}

\bibitem[{{Tanimoto} {et~al.}(2020){Tanimoto}, {Ueda}, {Odaka}, {Ogawa},
  {Yamada}, {Kawaguchi}, \& {Ichikawa}}]{Tanimoto_2020}
{Tanimoto}, A., {Ueda}, Y., {Odaka}, H., {et~al.} 2020, \apj, 897, 2,
  \dodoi{10.3847/1538-4357/ab96bc}

\bibitem[{{Tanimoto} {et~al.}(2022){Tanimoto}, {Ueda}, {Odaka}, {Yamada}, \&
  {Ricci}}]{Tanimoto2022}
{Tanimoto}, A., {Ueda}, Y., {Odaka}, H., {Yamada}, S., \& {Ricci}, C. 2022,
  \apjs, 260, 30, \dodoi{10.3847/1538-4365/ac5f59}

\bibitem[{{Toba} {et~al.}(2020){Toba}, {Yamada}, {Ueda}, {Ricci}, {Terashima},
  {Nagao}, {Wang}, {Tanimoto}, \& {Kawamuro}}]{Toba2020}
{Toba}, Y., {Yamada}, S., {Ueda}, Y., {et~al.} 2020, \apj, 888, 8,
  \dodoi{10.3847/1538-4357/ab5718}

\bibitem[{Tubín {et~al.}(2021)Tubín, Treister, D’Ago, Venturi, Bauer,
  Privon, Koss, Ricci, Comerford, \& Müller-Sánchez}]{Tubin_2021}
Tubín, D., Treister, E., D’Ago, G., {et~al.} 2021, The Astrophysical
  Journal, 911, 100, \dodoi{10.3847/1538-4357/abedba}

\bibitem[{{Uematsu} {et~al.}(2021){Uematsu}, {Ueda}, {Tanimoto}, {Kawamuro},
  {Setoguchi}, {Ogawa}, {Yamada}, \& {Odaka}}]{Uematsu2021}
{Uematsu}, R., {Ueda}, Y., {Tanimoto}, A., {et~al.} 2021, \apj, 913, 17,
  \dodoi{10.3847/1538-4357/abf0a2}

\bibitem[{{Urry} \& {Padovani}(1995)}]{Urry_1995}
{Urry}, C.~M., \& {Padovani}, P. 1995, \pasp, 107, 803, \dodoi{10.1086/133630}

\bibitem[{{Vasudevan} \& {Fabian}(2007)}]{Vasudevan_Fabian_2007}
{Vasudevan}, R.~V., \& {Fabian}, A.~C. 2007, \mnras, 381, 1235,
  \dodoi{10.1111/j.1365-2966.2007.12328.x}

\bibitem[{Weisskopf {et~al.}(2002)Weisskopf, Brinkman, Canizares, Garmire,
  Murray, \& Van~Speybroeck}]{Weisskopf_2002}
Weisskopf, M.~C., Brinkman, B., Canizares, C., {et~al.} 2002, Publications of
  the Astronomical Society of the Pacific, 114, 1–24, \dodoi{10.1086/338108}

\bibitem[{{Willingale} {et~al.}(2013){Willingale}, {Starling}, {Beardmore},
  {Tanvir}, \& {O'Brien}}]{Willingale_2013}
{Willingale}, R., {Starling}, R.~L.~C., {Beardmore}, A.~P., {Tanvir}, N.~R., \&
  {O'Brien}, P.~T. 2013, \mnras, 431, 394, \dodoi{10.1093/mnras/stt175}

\bibitem[{Yamada {et~al.}(2018)Yamada, Ueda, Oda, Tanimoto, Imanishi,
  Terashima, \& Ricci}]{Yamada_2018}
Yamada, S., Ueda, Y., Oda, S., {et~al.} 2018, The Astrophysical Journal, 858,
  106, \dodoi{10.3847/1538-4357/aabacb}

\bibitem[{{Yamada} {et~al.}(2021){Yamada}, {Ueda}, {Tanimoto}, {Imanishi},
  {Toba}, {Ricci}, \& {Privon}}]{Yamada_2021}
{Yamada}, S., {Ueda}, Y., {Tanimoto}, A., {et~al.} 2021, \apjs, 257, 61,
  \dodoi{10.3847/1538-4365/ac17f5}

\bibitem[{{Yamada} {et~al.}(2019){Yamada}, {Ueda}, {Tanimoto}, {Kawamuro},
  {Imanishi}, \& {Toba}}]{Yamada_2019}
---. 2019, \apj, 876, 96, \dodoi{10.3847/1538-4357/ab14f0}

\bibitem[{{Yamada} {et~al.}(2020){Yamada}, {Ueda}, {Tanimoto}, {Oda},
  {Imanishi}, {Toba}, \& {Ricci}}]{Yamada_2020}
---. 2020, \apj, 897, 107, \dodoi{10.3847/1538-4357/ab94b1}

\end{thebibliography}
\bibliographystyle{aasjournal}



\end{document}